\documentclass[%
 reprint,
superscriptaddress,
longbibliography
 amsmath,amssymb,
 aps,
 prl,
]{revtex4-2}

\usepackage{ulem}
\usepackage{soul}
\usepackage{gensymb}
\usepackage{textcomp}
\usepackage{hyperref}
\usepackage{graphicx}
\usepackage{dcolumn}
\usepackage{bm}
\usepackage{xcolor}

\newcommand{\LuNbSn}{LuNb$_6$Sn$_6$}
\newcommand{\ScVSn}{ScV$_6$Sn$_6$}
\newcommand{\HfFeGe}{HfFe$_6$Ge$_6$}

\begin{document}

\preprint{APS/123-QED}

\title{Pressure suppresses the density wave order in kagome metal LuNb$_6$Sn$_6$}

\author{William R. Meier}
\email{javamocham@gmail.com}
\affiliation{Materials Science \& Engineering Department, University of Tennessee Knoxville, Knoxville, TN 37996, USA}%

\author{David E. Graf}
\affiliation{National High Magnetic Field Laboratory, Tallahassee, FL, 32310, USA}%

\author{Brenden R. Ortiz}
\email{ortizbr@ornl.gov}
\affiliation{Materials Science \& Technology Division, Oak Ridge National Laboratory, Oak Ridge, TN 37831, USA}

\author{Shirin Mozaffari}
\affiliation{Materials Science \& Engineering Department, University of Tennessee Knoxville, Knoxville, TN 37996, USA}%

\author{David Mandrus}
\email{dmandrus@utk.edu}
\affiliation{Materials Science \& Engineering Department, University of Tennessee Knoxville, Knoxville, TN 37996, USA}%
\affiliation{Department of Physics \& Astronomy, University of Tennessee Knoxville, Knoxville, TN 37996, USA}%
\affiliation{Materials Science \& Technology Division, Oak Ridge National Laboratory, Oak Ridge, TN 37831, USA}%

\date{\today}

\begin{abstract}

\begin{center}
Dancing tins pair up,%

But compressing the framework%

Thwarts the displacements.%
\end{center}

The density waves that develop in kagome metals \ScVSn{} and \LuNbSn{} at low temperature appear to arise from under-filled atomic columns within a V-Sn or Nb-Sn scaffolding.
Compressing this network with applied pressure in \ScVSn{} suppressed the structural transition temperature by constraining atomic rattling and inhibiting the shifts that define the structural modulation.
We predicted that the density wave transition in \LuNbSn{} at 68\,K would be suppressed by pressure as well.
In this brief study, we examine the pressure dependence of the density wave transition by measuring resistance vs temperature up to 2.26\,GPa.
We found the transition temperature is smoothly depressed and disappears around 1.9\,GPa.
This result not only addresses our prediction, but strengthens the rattling chains origin of structural instabilities in the \HfFeGe-type kagome metals.

\end{abstract}

\maketitle


\section{Introduction}
\label{sec:Intro}

\begin{figure*}
    \includegraphics[width=17.2cm]{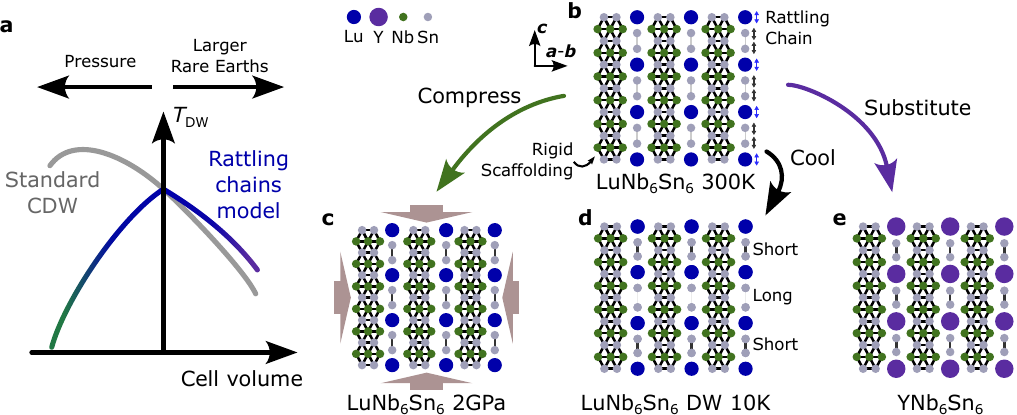}
    \caption{\label{fig:RattlingIntro} 
    	\textbf{(a)} The transition temperature of traditional charge density wave (CDW) systems is generally expected to evolve smoothly with unit cell volume.\cite{Sacchetti2007_PressureDepCDWGap-RTe3,Luccas2015_CDW-La1-xCexSb2,Budko2022_PressureSuppressCDW-LaSb2,Budko2006_Pressure+ChemSub-CDW_LaAgSb2}  
        In contrast, density wave (DW) orders that develop due to under-filled rattling chains like those in \ScVSn{} and \LuNbSn{} are expected to be suppressed by lattice expansion, by rare earth doping, and by compression under pressure.
        \textbf{(b)} Cartoon structure of \LuNbSn{} along the (110) plane. 
        A rigid scaffolding of Nb and Sn atom (green and gray) surround under-filled channels hosting a Lu-Sn-Sn chains. 
        The extra space provided by the small Lu atoms allows the chains to rattle and displace, forming a DW structural modulation at low temperature \textbf{(d)}. 
        Applying pressure \textbf{(c)} and swapping Lu with larger rare earths \textbf{(e)} inhibits both rattling and DW formation.
    }	
\end{figure*}

Researchers have been enamored by the kagome lattice for the last decades due to the exciting behaviors that can arise in materials hosting this structural motif. 
In metallic systems, the connectivity of atoms on kagome sites generates electronic structures with desirable states like flat bands, van Hove singularities, and Dirac nodes.\cite{Wang2013_KagomeCompetingOrders,Kiesel2013_FermiSurfaceInstabilites-KagomeHubbard}
Among metallic kagome compounds, structural modulations have emerged as a reoccurring theme. The CsV$_3$Sb$_5$-family,\cite{Ortiz2019_DiscoveryAV3Sb5,Wilson2024_AV3Sb5-Review} CsCr$_3$Sb$_5$,\cite{Li2025_CsCr3Sb5-ElectronicStructure+FlatBands} V$_3$Sb$_2$,\cite{Wang2022_DensityWave-V3Sb2} La(Ru,Fe)$_3$Si$_2$,\cite{Plokhikh2024_CDW-LuRu1-xFex3Si2} Yb$_{0.5}$Co$_3$Ge$_3$,\cite{Wang2022_Properties+PhaseTransitionYb0.5Co3Ge3} hexagonal-FeGe,\cite{Teng2023_Mag+CDW-inHexFeGe,Teng2022_CDWinHexFeGe,Chen2024_LongRangeCDW+DimerizationHexFeGe} and \ScVSn{} \cite{Arachchige2022_CDW-ScV6Sn6} all develop periodic structural distortions on cooling. 

The density wave instability observed below 92\,K in \ScVSn{} has received significant attention due to some of its peculiar features.
These include curious transport properties,\cite{Mozaffari2024_RV6Sn6-SublinearResistivity,DeStefano2023_ScV6Sn6-TransportRevealedByMagTrasport,Yi2024_QuantumOscillationsScV6Sn6,Kuo2024_ScV6Sn6-Termoelectric} 
competing density wave instabilities,\cite{Cao2023_CompetingCDWs-ScV6Sn6,Korshunov2023_SofteningFlatPhononScV6Sn6,Hu2025_FlatPhononSoftModes-ScV6Sn6-PhononCalcs,Pokharel2023_FrustratedChargeOrder+CooperativeDistortions-ScV6Sn6,Subedi2024_OrderByDisorderCDW-ScV6Sn6,Wang2024_OriginOfCompetingCDWs-ScV6Sn6,Tan2023_AbundantLatticeInstabilites-ScV6Sn6}
and subtle band structure reconstruction.\cite{Tuniz2023_Dynamics+ResilienceOfCDW-ScV6Sn6,Lee2024_NatureOfCDW-ScV6Sn6,Kim2023_IR-ProbeCDWGap-ScV6Sn6,Cheng2024_UntangleCDWBulk-Surface-ScV6Sn6,Yang2024_UnveilingCDWMechanism-ScV6Sn6,Kundu2024_LowEnergyElectronicStructure-ScV6Sn6}
The structural modulation that develops in \ScVSn{} has characteristics that distinguish it from traditional charge density waves (CDW).
In particular, the transition temperature can be suppressed by compressing the lattice with pressure \cite{Zhang2022_ScV6Sn6-Pressure} and by filling the lattice by larger rare earth elements.\cite{Meier2023_TinyScAllowChainsToRattle-Lu+YDopedScV6Sn6} 

This is different from usual behavior expected for the lattice volume dependence for CDW systems.
Expanding the lattice through isovalent doping and shrinking it with pressure are expected to have opposite impacts on the Fermi-surface details that underlie CDW formation.
Therefore, these system often have common trends of transition temperature with lattice volume with both doping and pressure (Fig.~\ref{fig:RattlingIntro}a).\cite{Sacchetti2007_PressureDepCDWGap-RTe3,Luccas2015_CDW-La1-xCexSb2,Budko2022_PressureSuppressCDW-LaSb2,Budko2006_Pressure+ChemSub-CDW_LaAgSb2} 

In fact, there is direct evidence that the modulation in \ScVSn{} does not arise from the Fermi surface nesting common for CDWs.\cite{Tan2023_AbundantLatticeInstabilites-ScV6Sn6,Cheng2024_NanoscaleVisualizationOfCDW-ScV6Sn6,Hu2024_PhononPromotedCDW-ScV6Sn6,Cao2023_CompetingCDWs-ScV6Sn6,Korshunov2023_SofteningFlatPhononScV6Sn6,Hu2025_FlatPhononSoftModes-ScV6Sn6-PhononCalcs,Yang2024_UnveilingCDWMechanism-ScV6Sn6}
This motivated the development of a rattling model to understand the structural instability in the material.\cite{Meier2023_TinyScAllowChainsToRattle-Lu+YDopedScV6Sn6}
In short, \HfFeGe-type $AM_6X_6$ compounds can have rattling instabilities whenever a small $A$ atom sits in a sufficiently large $M_6X_6$ scaffolding. 
In \ScVSn{} this extra space leads to the formation of Sn-Sn bond modulation that characterize the low-temperature structure.\cite{Korshunov2023_SofteningFlatPhononScV6Sn6,Hu2024_PhononPromotedCDW-ScV6Sn6}

We recently demonstrated that this rattling instability model extends to the extensive $Ln$Nb$_6$Sn$_6$ family of compounds.\cite{Ortiz2025_166StabilityFrontiers+LnNb6Sn6-Family+DW-LuNb6Sn6}
In the compounds with the smallest rare earths, x-ray experiments reveal rattling of rare earth-Sn chains within the Nb-Sn network (Fig.~\ref{fig:RattlingIntro}b).
Crucially, \LuNbSn{} develops a density wave (DW) order at approximately 68\,K with static Sn-Sn bond modulation (Fig.~\ref{fig:RattlingIntro}d) analogous to that in \ScVSn{}. 
We demonstrated that replacing lutetium with larger rare earth elements not only inhibits the rattling at high temperatures, but also prevents the development of a DW (Fig.~\ref{fig:RattlingIntro}e).

This observation reinforced the rattling chain model developed for \ScVSn\cite{Meier2023_TinyScAllowChainsToRattle-Lu+YDopedScV6Sn6} and inspired an important prediction: \textbf{We hypothesized that the density wave transition temperature of \LuNbSn{} will be suppressed with pressure (Fig.~\ref{fig:RattlingIntro}a).} 
We anticipate that rattling and Sn-Sn bond modulation will be inhibited as the Nb-Sn framework is compressed as seen in \ScVSn (Fig.~\ref{fig:RattlingIntro}c).

In this paper we present an expedient test of this idea by measuring the temperature-dependent resistance of \LuNbSn{} under pressure to 2.26\,GPa. 
We clearly observe that the transition temperature is suppressed as pressure is increased. 
The density wave in \LuNbSn{} is suppressed by applied pressure, confirming our prediction.
We further compare the pressure evolution of the density wave transition in \ScVSn{} and Lu-doped \ScVSn{} reported previously.
Our work reaffirms the applicability of the rattling chain model in under-filled network compounds like the  \HfFeGe-type materials. 
 
\section{Methods}
\label{sec:Methods}

Single crystal growth of LuNb$_6$Sn$_6$ was performed using the self-flux technique.\cite{Ortiz2025_166StabilityFrontiers+LnNb6Sn6-Family+DW-LuNb6Sn6} 
Elemental reagents were combined in a ratio of 8:2:90 = Lu:Nb:Sn. 
We utilized Ames Lab Lu metal, Nb powder (Alfa, 99.9\%), and Sn shot (Alfa, 99.9\%). 
Reagents were placed into 5\,mL Al$_2$O$_3$ (Canfield) crucibles fitted with a catch crucible/porous frit and sealed within fused quartz ampoules with approximately 0.6--0.7\,atm of argon cover gas.\cite{Canfield2016_CanfieldCrucibleSets} 
Samples were heated to 1150\,\degree C and thermalized for 18\,h before cooling to 900\,\degree C at a rate of 1\,\degree C/hr. 
Samples are subsequently centrifuged at 900\,\degree C. 
Single crystals are small, well-faceted, gray-metallic, hexagonal plates and blocks. 
The samples are stable in air, water, and common solvents. 

For these high-pressure measurements, a 4\,mm ID hybrid cell was used with a BeCu outer wall and NiCrAl alloy for an insert. 
Copper wires and a fiber optic were fixed in place through the sample platform with Stycast 2850 FT epoxy. 
A hexagonal block-shaped crystal of \LuNbSn{} was hand polished into a minuscule bar with a 80$\times$120\,$\mu$m cross section using sand paper. 
Contacts to the sample were made with 0.0005\,in platinum wire in a conventional 4-contact arrangement held in place with Epotek H20E silver epoxy cured for 30\,min at 130\,\degree C. 
The pressure medium is Daphne 7575 oil which has been shown to remain hydrostatic to pressures above 3\,GPa.\cite{Stasko2020_PressureMediaDaphneOil7000}
Transport measurements of the sample were made with a Lake Shore 372 resistance bridge with 3708 pre-amp/scanner using 1\,mA and a 5\,s averaging time. 
All data was collected with cooling curves taken at a rate of 0.4\,K/min. 
Accurate temperatures were ensured by fixing a Cernox thermometer to the outside of the pressure cell and allowing for additional exchange gas in the sample space of the PPMS where the experiment took place.
Resistance data was averaged over 15 measurements to reduce the noise in the derivative. Raw data is provided in the supplemental materials.\cite{LuNb6Sn6Pressure_SupplementaryData}

The fiber optics ends were cleaved cleanly to allow for a small chip of ruby to be glued to the end inside of the sample volume. 
Green laser light (532\,nm) was transmitted to the ruby chip and the return fluorescence spectrum was determined by an Ocean Optics spectrometer. 
Ruby peaks are known to shift at a rate of 0.365\,nm/GPa\cite{Piermarini1975_RubyFluorescenceTo195kbar} and a room temperature and low temperature value for pressure were recorded for each curve. 
The reported pressures are from the region of interest (low temperatures) where the charge density wave was suppressed.

\section{Results}
\label{sec:Results}

\begin{figure}
    \includegraphics[width=8.6cm]{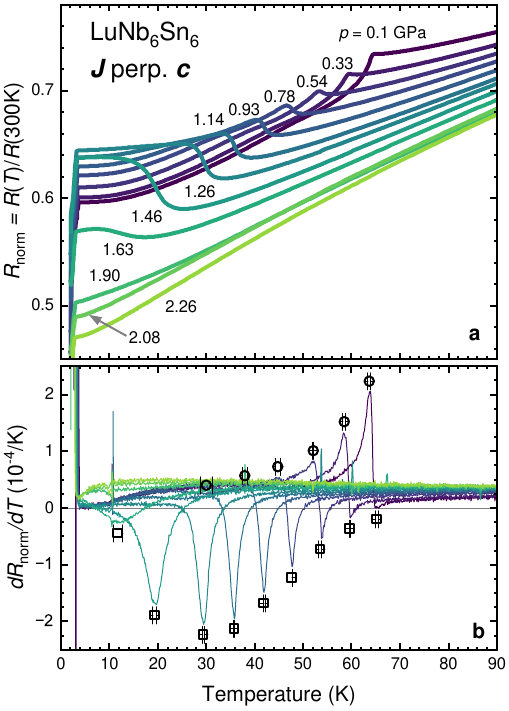}
    \caption{\label{fig:RT_Plot} 
    	\textbf{(a)} normalized resistance vs temperature data for LuNb$_6$Sn$_6$ at increasing pressures measured on cooling. 
        The sharp drops near 3\,K are from a tin impurity.
        \textbf{(b)} temperature derivatives of resistivity emphasize transition features.
        Maxima and Minima of the derivative are marked with circles and squares, respectively. 
        Error bars communicate estimated temperature uncertainty.
    }	
\end{figure}

Figure~\ref{fig:RT_Plot}a presents the normalized resistance vs temperature data obtained for a series of pressure.
The labeled pressures were estimated at base temperature.
At the lowest pressure, resistivity drops sharply on cooling through 64\,K. 
This corresponds to the development of the density wave modulation observed by x-ray diffraction at zero pressure at 68\,K.\cite{Ortiz2025_166StabilityFrontiers+LnNb6Sn6-Family+DW-LuNb6Sn6}
In \ScVSn{} this drop has been tied to an increase in carrier mobility in the density wave phase.\cite{Arachchige2022_CDW-ScV6Sn6,Hu2023_OpticalConductivity+BandStructure-ScV6Sn6}

The shape of the resistance curve evolves neatly with increasing pressure (lighter colors).
The cliff-like drop at 0.1\,GPa evolves into a cusp-shaped peak at 0.54 and 0.78\,GPa. This further morphs into a broader step-up on cooling exemplified by the 1.26\,GPa curve.
The pressure evolution of this resistance feature mimics that observed in \ScVSn\cite{Zhang2022_ScV6Sn6-Pressure} and (Sc$_{0.944}$Lu$_{0.054}$)V$_6$Sn$_6$.\cite{Meier2023_TinyScAllowChainsToRattle-Lu+YDopedScV6Sn6}
Importantly, the high-pressure jump-up feature is most dramatic in \LuNbSn.

In Fig.~\ref{fig:RT_Plot}a there is a sharp drop in resistance between 2.8 and 3.7\,K. 
A linear fit to the onset temperature vs pressure yields a slope of -0.41$\pm$0.01\,K/GPa with an intercept of 3.74$\pm$0.02 K. 
This agrees exquisitely with the pressure dependence the superconducting critical temperature of tin metal ($T_\mathrm{c} = 3.733 - (0.43\,\mathrm{K/GPa})\,p$).\cite{Brandt1965_SuperconductingMetalsUnderPressure}
Despite careful polishing, there is a Sn inclusion in our sample.

The evolving nature of the density wave transition feature in this data set necessitates a systematic approach to extracting the transition temperatures.
We plot the temperature derivative of resistance presented in Fig.~\ref{fig:RT_Plot}b to assist us.
We selected the maxima and minima in $\frac{dR_{\mathrm{norm}}}{dT}(T)$ as characteristic temperatures as they correspond to the positive and negative slope extremes of $R_{\mathrm{norm}}(T)$.
We identify these temperatures with estimated uncertainties with circles and squares in Fig.~\ref{fig:RT_Plot}b.

At low pressures where a step-down feature is observed, the maximum of $\frac{dR_{\mathrm{norm}}}{dT}(T)$ is best descriptor of this first-order transition temperature.
In contrast, at higher pressure (like 1.26\,GPa) the steepest negative slope of $R_{\mathrm{norm}}(T)$ appears to be the best descriptor for a broadened first-order transition.
At intermediate pressures, we could imagine the transition temperature at the cusp in which naturally lies between the maxima and minima of $\frac{dR_{\mathrm{norm}}}{dT}(T)$.

Importantly, we see no evidence of bulk superconductivity emerging as the density wave disappears.
This was also true for \ScVSn{} under pressure,\cite{Zhang2022_ScV6Sn6-Pressure} doped \ScVSn, and Lu-doped \ScVSn{} under pressure.\cite{Lee2024_NatureOfCDW-ScV6Sn6,Meier2023_TinyScAllowChainsToRattle-Lu+YDopedScV6Sn6}
The first order nature of this transition in these compounds may provide a poor starting point for building a superconducting state.

\begin{figure}
    \includegraphics[width=8.6cm]{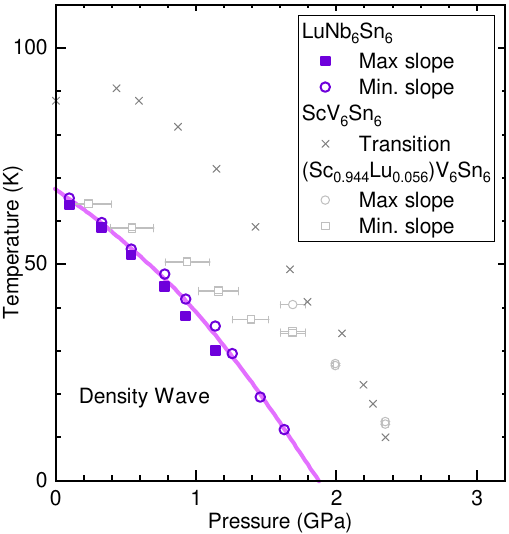}
    \caption{\label{fig:PhaseDiagram} 
    	Phase diagram reveals the suppression of the phase transition in \LuNbSn{} with pressure.
        Transition temperatures from \ScVSn\cite{Zhang2022_ScV6Sn6-Pressure} and Lu-doped \ScVSn\cite{Meier2023_TinyScAllowChainsToRattle-Lu+YDopedScV6Sn6} are included, showing higher critical pressures.
    }	
\end{figure}

Figure~\ref{fig:PhaseDiagram} presents the phase diagram of \LuNbSn{} under pressure.
The purple data clearly reveal that the transition temperature is smoothly suppressed with applied pressure and disappears by roughly 1.9\,GPa.
The solid line is included to emphasize the trend.

\section{Discussion}
\label{sec:Discussion}

We have confirmed our prediction that the density wave phase in \LuNbSn{} would be suppressed by pressure.
This supports the rattling chains picture developed for \ScVSn\cite{Meier2023_TinyScAllowChainsToRattle-Lu+YDopedScV6Sn6} we extended to \LuNbSn.\cite{Ortiz2025_166StabilityFrontiers+LnNb6Sn6-Family+DW-LuNb6Sn6} 
Compressing the Nb-Sn scaffolding removes the extra space in the structure, inhibiting DW formation. 

It is also possible that compressing the lattice also modifies electronic structure in a way that could disfavor DW development. 
We suspect these electronic changes are less important based on numerous studies finding that Fermi surface details are not the key driver of DW order in \ScVSn.\cite{Tan2023_AbundantLatticeInstabilites-ScV6Sn6,Cheng2024_NanoscaleVisualizationOfCDW-ScV6Sn6,Hu2024_PhononPromotedCDW-ScV6Sn6,Cao2023_CompetingCDWs-ScV6Sn6,Korshunov2023_SofteningFlatPhononScV6Sn6,Hu2025_FlatPhononSoftModes-ScV6Sn6-PhononCalcs,Yang2024_UnveilingCDWMechanism-ScV6Sn6}
To address this possibility, we propose studying Sc substituted \LuNbSn{} to further examine the importance of steric effects in DW development. 
The rattling chains picture would predict that replacing Lu with tiny Sc would exacerbate the under-filling of the scaffolding increasing the DW transitions temperature.

Demonstrating broader applicability of the rattling chain model beyond the $R$V$_{6}$Sn$_{6}$ family strengthens its value for materials design.
Displacement fluctuations can be engineered by under-filling channels in compounds with rigid scaffolding networks. 
This could enhance thermoelectric performance and induce density wave transitions. 
In addition, identifying the critical pressure required to prevent density wave formation will aid computational efforts exploring the criteria for structural instabilities in the \HfFeGe -type metals.

The pressure evolution of the transition temperatures in \ScVSn{} and (Sc$_{0.944}$Lu$_{0.056}$)V$_6$Sn$_6$ are plotted fig.~\ref{fig:PhaseDiagram}.\cite{Zhang2022_ScV6Sn6-Pressure}
Curiously, although the Lu-doped material has a significantly reduced transition temperature at zero pressure, it shares a comparable critical pressure with \ScVSn.
\LuNbSn{} has a notably lower critical pressure.
This difference might be a result of the Nb-Sn framework being more compressible than the V-Sn.
This should be addressable with DFT calculations or diffraction experiments under pressure.

Next we turn to a trend common to all three materials, the evolution of the shape of the resistance transition signature with pressure.
In most density wave phases the resistance is higher in the low temperature phase due to a loss of carrier when band reconstruct.\cite{Gruner1994_DensityWavesInSolids}
In \ScVSn, only the Sn1 p-bands show significant reconstruction accounting for a minor reduction in carrier concentration.\cite{Tan2023_AbundantLatticeInstabilites-ScV6Sn6,Tuniz2023_Dynamics+ResilienceOfCDW-ScV6Sn6,Hu2025_FlatPhononSoftModes-ScV6Sn6-PhononCalcs,Liu2024_DrivingMechanism+FluctuationsOfCDWs-ScV6Sn6}
It was proposed that the reduced resistance in the DW phase is due to sudden loss of strong scattering\cite{Arachchige2022_CDW-ScV6Sn6,Mozaffari2024_RV6Sn6-SublinearResistivity,Hu2023_OpticalConductivity+BandStructure-ScV6Sn6} by low energy phonons associated with the rattling chains.\cite{Cao2023_CompetingCDWs-ScV6Sn6,Korshunov2023_SofteningFlatPhononScV6Sn6,Pokharel2023_FrustratedChargeOrder+CooperativeDistortions-ScV6Sn6}
It is likely the same explanation lies behind the step down transition feature observed in \LuNbSn{} at zero-pressure.

Why then does the transition step difference change sign on increasing pressure in both \ScVSn{} and \LuNbSn?
First, this might represent a crossover in the relative impact of the change of carrier concentration and mobility as the density wave develops.
Alternatively, there may be a change in the nature of the density wave order at higher pressures.
For example, applying pressure might favor the competing 1/3 1/3 1/2 density wave\cite{Cao2023_CompetingCDWs-ScV6Sn6,Korshunov2023_SofteningFlatPhononScV6Sn6,Pokharel2023_FrustratedChargeOrder+CooperativeDistortions-ScV6Sn6} over the 1/3 1/3 1/3 observed at zero-pressure in both compounds.\cite{Arachchige2022_CDW-ScV6Sn6,Ortiz2025_166StabilityFrontiers+LnNb6Sn6-Family+DW-LuNb6Sn6}
These ideas could be tested by low-temperature Raman spectroscopy, optical conductivity, or x-ray scattering under pressure.

\section{Conclusion}

The density wave instabilities of \ScVSn{} and \LuNbSn{} result from the extra space provided by undersized atoms in a rigid V-Sn or Nb-Sn network.
In the scandium compound, compressing this network with pressure inhibits the displacements that constitute the density wave, reducing the transition temperature.
We predicted that the transition should be suppressed with pressure in \LuNbSn{} as well.
Our temperature dependent resistance measurements reveal confirm this.
This observation emphasizes the value of the rattling chain model for understanding the tunability of density wave instabilities in filled network compounds.

\section{Acknowledgments}
\begin{acknowledgments}
	\label{sec:Acknowledgment}
	
	WRM acknowledges support from the Gordon and Betty Moore Foundation’s EPiQS Initiative, Grant GBMF9069 awarded to D.M..
    Work by B.R.O. was supported by the U.S. Department of Energy, Office of Science, Basic Energy Sciences, Materials Sciences and Engineering Division.
    S.M. and D.M. acknowledge the support from AFOSR MURI (Novel Light-Matter Interactions in Topologically Non-Trivial Weyl Semimetal Structures and Systems), Grant No. FA9550-20-1-0322. 
    A portion of this work was performed at the National High Magnetic Field Laboratory, which is supported by National Science Foundation Cooperative Agreement No. DMR-2128556 and the State of Florida.  
	
\end{acknowledgments}


\begin{thebibliography}{46}%
\makeatletter
\providecommand \@ifxundefined [1]{%
 \@ifx{#1\undefined}
}%
\providecommand \@ifnum [1]{%
 \ifnum #1\expandafter \@firstoftwo
 \else \expandafter \@secondoftwo
 \fi
}%
\providecommand \@ifx [1]{%
 \ifx #1\expandafter \@firstoftwo
 \else \expandafter \@secondoftwo
 \fi
}%
\providecommand \natexlab [1]{#1}%
\providecommand \enquote  [1]{``#1''}%
\providecommand \bibnamefont  [1]{#1}%
\providecommand \bibfnamefont [1]{#1}%
\providecommand \citenamefont [1]{#1}%
\providecommand \href@noop [0]{\@secondoftwo}%
\providecommand \href [0]{\begingroup \@sanitize@url \@href}%
\providecommand \@href[1]{\@@startlink{#1}\@@href}%
\providecommand \@@href[1]{\endgroup#1\@@endlink}%
\providecommand \@sanitize@url [0]{\catcode `\\12\catcode `\$12\catcode `\&12\catcode `\#12\catcode `\^12\catcode `\_12\catcode `\%12\relax}%
\providecommand \@@startlink[1]{}%
\providecommand \@@endlink[0]{}%
\providecommand \url  [0]{\begingroup\@sanitize@url \@url }%
\providecommand \@url [1]{\endgroup\@href {#1}{\urlprefix }}%
\providecommand \urlprefix  [0]{URL }%
\providecommand \Eprint [0]{\href }%
\providecommand \doibase [0]{https://doi.org/}%
\providecommand \selectlanguage [0]{\@gobble}%
\providecommand \bibinfo  [0]{\@secondoftwo}%
\providecommand \bibfield  [0]{\@secondoftwo}%
\providecommand \translation [1]{[#1]}%
\providecommand \BibitemOpen [0]{}%
\providecommand \bibitemStop [0]{}%
\providecommand \bibitemNoStop [0]{.\EOS\space}%
\providecommand \EOS [0]{\spacefactor3000\relax}%
\providecommand \BibitemShut  [1]{\csname bibitem#1\endcsname}%
\let\auto@bib@innerbib\@empty
\bibitem [{\citenamefont {Sacchetti}\ \emph {et~al.}(2007)\citenamefont {Sacchetti}, \citenamefont {Arcangeletti}, \citenamefont {Perucchi}, \citenamefont {Baldassarre}, \citenamefont {Postorino}, \citenamefont {Lupi}, \citenamefont {Ru}, \citenamefont {Fisher},\ and\ \citenamefont {Degiorgi}}]{Sacchetti2007_PressureDepCDWGap-RTe3}%
  \BibitemOpen
  \bibfield  {author} {\bibinfo {author} {\bibfnamefont {A.}~\bibnamefont {Sacchetti}}, \bibinfo {author} {\bibfnamefont {E.}~\bibnamefont {Arcangeletti}}, \bibinfo {author} {\bibfnamefont {A.}~\bibnamefont {Perucchi}}, \bibinfo {author} {\bibfnamefont {L.}~\bibnamefont {Baldassarre}}, \bibinfo {author} {\bibfnamefont {P.}~\bibnamefont {Postorino}}, \bibinfo {author} {\bibfnamefont {S.}~\bibnamefont {Lupi}}, \bibinfo {author} {\bibfnamefont {N.}~\bibnamefont {Ru}}, \bibinfo {author} {\bibfnamefont {I.~R.}\ \bibnamefont {Fisher}},\ and\ \bibinfo {author} {\bibfnamefont {L.}~\bibnamefont {Degiorgi}},\ }\bibfield  {title} {\bibinfo {title} {Pressure dependence of the charge-density-wave gap in rare-earth tritellurides},\ }\href {https://doi.org/10.1103/physrevlett.98.026401} {\bibfield  {journal} {\bibinfo  {journal} {Physical Review Letters}\ }\textbf {\bibinfo {volume} {98}},\ \bibinfo {pages} {026401} (\bibinfo {year} {2007})}\BibitemShut {NoStop}%
\bibitem [{\citenamefont {Luccas}\ \emph {et~al.}(2015)\citenamefont {Luccas}, \citenamefont {Fente}, \citenamefont {Hanko}, \citenamefont {Correa-Orellana}, \citenamefont {Herrera}, \citenamefont {Climent-Pascual}, \citenamefont {Azpeitia}, \citenamefont {P\'erez-Casta\~neda}, \citenamefont {Osorio}, \citenamefont {Salas-Colera}, \citenamefont {Nemes}, \citenamefont {Mompean}, \citenamefont {Garc\'ia-Hern\'andez}, \citenamefont {Rodrigo}, \citenamefont {Ramos}, \citenamefont {Guillam\'on}, \citenamefont {Vieira},\ and\ \citenamefont {Suderow}}]{Luccas2015_CDW-La1-xCexSb2}%
  \BibitemOpen
  \bibfield  {author} {\bibinfo {author} {\bibfnamefont {R.~F.}\ \bibnamefont {Luccas}}, \bibinfo {author} {\bibfnamefont {A.}~\bibnamefont {Fente}}, \bibinfo {author} {\bibfnamefont {J.}~\bibnamefont {Hanko}}, \bibinfo {author} {\bibfnamefont {A.}~\bibnamefont {Correa-Orellana}}, \bibinfo {author} {\bibfnamefont {E.}~\bibnamefont {Herrera}}, \bibinfo {author} {\bibfnamefont {E.}~\bibnamefont {Climent-Pascual}}, \bibinfo {author} {\bibfnamefont {J.}~\bibnamefont {Azpeitia}}, \bibinfo {author} {\bibfnamefont {T.}~\bibnamefont {P\'erez-Casta\~neda}}, \bibinfo {author} {\bibfnamefont {M.~R.}\ \bibnamefont {Osorio}}, \bibinfo {author} {\bibfnamefont {E.}~\bibnamefont {Salas-Colera}}, \bibinfo {author} {\bibfnamefont {N.~M.}\ \bibnamefont {Nemes}}, \bibinfo {author} {\bibfnamefont {F.~J.}\ \bibnamefont {Mompean}}, \bibinfo {author} {\bibfnamefont {M.}~\bibnamefont {Garc\'ia-Hern\'andez}}, \bibinfo {author} {\bibfnamefont {J.~G.}\ \bibnamefont {Rodrigo}}, \bibinfo {author} {\bibfnamefont {M.~A.}\ \bibnamefont
  {Ramos}}, \bibinfo {author} {\bibfnamefont {I.}~\bibnamefont {Guillam\'on}}, \bibinfo {author} {\bibfnamefont {S.}~\bibnamefont {Vieira}},\ and\ \bibinfo {author} {\bibfnamefont {H.}~\bibnamefont {Suderow}},\ }\bibfield  {title} {\bibinfo {title} {Charge density wave in layered {La}$_{1-x}${Ce}$_{x}${Sb}$_{2}$},\ }\href {https://doi.org/10.1103/physrevb.92.235153} {\bibfield  {journal} {\bibinfo  {journal} {Physical Review B}\ }\textbf {\bibinfo {volume} {92}},\ \bibinfo {pages} {235153} (\bibinfo {year} {2015})}\BibitemShut {NoStop}%
\bibitem [{\citenamefont {Bud’ko}\ \emph {et~al.}({\natexlab{a}})\citenamefont {Bud’ko}, \citenamefont {Huyan}, \citenamefont {Herrera-Siklody},\ and\ \citenamefont {Canfield}}]{Budko2022_PressureSuppressCDW-LaSb2}%
  \BibitemOpen
  \bibfield  {author} {\bibinfo {author} {\bibfnamefont {S.~L.}\ \bibnamefont {Bud’ko}}, \bibinfo {author} {\bibfnamefont {S.}~\bibnamefont {Huyan}}, \bibinfo {author} {\bibfnamefont {P.}~\bibnamefont {Herrera-Siklody}},\ and\ \bibinfo {author} {\bibfnamefont {P.~C.}\ \bibnamefont {Canfield}},\ }\bibfield  {title} {\bibinfo {title} {Rapid suppression of charge density wave transition in {LaSb}$_{2}$ under pressure},\ }\href {https://doi.org/10.1080/14786435.2022.2159561} {\bibfield  {journal} {\bibinfo  {journal} {Philosophical Magazine}\ }\textbf {\bibinfo {volume} {103}},\ \bibinfo {pages} {561} ({\natexlab{a}})}\BibitemShut {NoStop}%
\bibitem [{\citenamefont {Bud’ko}\ \emph {et~al.}({\natexlab{b}})\citenamefont {Bud’ko}, \citenamefont {Wiener}, \citenamefont {Ribeiro}, \citenamefont {Canfield}, \citenamefont {Lee}, \citenamefont {Vogt},\ and\ \citenamefont {Lacerda}}]{Budko2006_Pressure+ChemSub-CDW_LaAgSb2}%
  \BibitemOpen
  \bibfield  {author} {\bibinfo {author} {\bibfnamefont {S.~L.}\ \bibnamefont {Bud’ko}}, \bibinfo {author} {\bibfnamefont {T.~A.}\ \bibnamefont {Wiener}}, \bibinfo {author} {\bibfnamefont {R.~A.}\ \bibnamefont {Ribeiro}}, \bibinfo {author} {\bibfnamefont {P.~C.}\ \bibnamefont {Canfield}}, \bibinfo {author} {\bibfnamefont {Y.}~\bibnamefont {Lee}}, \bibinfo {author} {\bibfnamefont {T.}~\bibnamefont {Vogt}},\ and\ \bibinfo {author} {\bibfnamefont {A.~H.}\ \bibnamefont {Lacerda}},\ }\bibfield  {title} {\bibinfo {title} {Effect of pressure and chemical substitutions on the charge-density-wave in {LaAgSb}$_{2}$},\ }\href {https://doi.org/10.1103/physrevb.73.184111} {\bibfield  {journal} {\bibinfo  {journal} {Physical Review B}\ }\textbf {\bibinfo {volume} {73}},\ \bibinfo {pages} {184111} ({\natexlab{b}})}\BibitemShut {NoStop}%
\bibitem [{\citenamefont {Wang}\ \emph {et~al.}(2013)\citenamefont {Wang}, \citenamefont {Li}, \citenamefont {Xiang},\ and\ \citenamefont {Wang}}]{Wang2013_KagomeCompetingOrders}%
  \BibitemOpen
  \bibfield  {author} {\bibinfo {author} {\bibfnamefont {W.-S.}\ \bibnamefont {Wang}}, \bibinfo {author} {\bibfnamefont {Z.-Z.}\ \bibnamefont {Li}}, \bibinfo {author} {\bibfnamefont {Y.-Y.}\ \bibnamefont {Xiang}},\ and\ \bibinfo {author} {\bibfnamefont {Q.-H.}\ \bibnamefont {Wang}},\ }\bibfield  {title} {\bibinfo {title} {{Competing electronic orders on kagome lattices at van Hove filling}},\ }\href {https://doi.org/10.1103/physrevb.87.115135} {\bibfield  {journal} {\bibinfo  {journal} {Physical Review B—Condensed Matter and Materials Physics}\ }\textbf {\bibinfo {volume} {87}},\ \bibinfo {pages} {115135} (\bibinfo {year} {2013})}\BibitemShut {NoStop}%
\bibitem [{\citenamefont {Kiesel}\ \emph {et~al.}(2013)\citenamefont {Kiesel}, \citenamefont {Platt},\ and\ \citenamefont {Thomale}}]{Kiesel2013_FermiSurfaceInstabilites-KagomeHubbard}%
  \BibitemOpen
  \bibfield  {author} {\bibinfo {author} {\bibfnamefont {M.~L.}\ \bibnamefont {Kiesel}}, \bibinfo {author} {\bibfnamefont {C.}~\bibnamefont {Platt}},\ and\ \bibinfo {author} {\bibfnamefont {R.}~\bibnamefont {Thomale}},\ }\bibfield  {title} {\bibinfo {title} {{Unconventional Fermi surface instabilities in the kagome Hubbard model}},\ }\href {https://doi.org/10.1103/physrevlett.110.126405} {\bibfield  {journal} {\bibinfo  {journal} {Physical Review Letters}\ }\textbf {\bibinfo {volume} {110}},\ \bibinfo {pages} {126405} (\bibinfo {year} {2013})}\BibitemShut {NoStop}%
\bibitem [{\citenamefont {Ortiz}\ \emph {et~al.}(2019)\citenamefont {Ortiz}, \citenamefont {Gomes}, \citenamefont {Morey}, \citenamefont {Winiarski}, \citenamefont {Bordelon}, \citenamefont {Mangum}, \citenamefont {Oswald}, \citenamefont {Rodriguez-Rivera}, \citenamefont {Neilson}, \citenamefont {Wilson}, \citenamefont {Ertekin}, \citenamefont {McQueen},\ and\ \citenamefont {Toberer}}]{Ortiz2019_DiscoveryAV3Sb5}%
  \BibitemOpen
  \bibfield  {author} {\bibinfo {author} {\bibfnamefont {B.~R.}\ \bibnamefont {Ortiz}}, \bibinfo {author} {\bibfnamefont {L.~C.}\ \bibnamefont {Gomes}}, \bibinfo {author} {\bibfnamefont {J.~R.}\ \bibnamefont {Morey}}, \bibinfo {author} {\bibfnamefont {M.}~\bibnamefont {Winiarski}}, \bibinfo {author} {\bibfnamefont {M.}~\bibnamefont {Bordelon}}, \bibinfo {author} {\bibfnamefont {J.~S.}\ \bibnamefont {Mangum}}, \bibinfo {author} {\bibfnamefont {I.~W.~H.}\ \bibnamefont {Oswald}}, \bibinfo {author} {\bibfnamefont {J.~A.}\ \bibnamefont {Rodriguez-Rivera}}, \bibinfo {author} {\bibfnamefont {J.~R.}\ \bibnamefont {Neilson}}, \bibinfo {author} {\bibfnamefont {S.~D.}\ \bibnamefont {Wilson}}, \bibinfo {author} {\bibfnamefont {E.}~\bibnamefont {Ertekin}}, \bibinfo {author} {\bibfnamefont {T.~M.}\ \bibnamefont {McQueen}},\ and\ \bibinfo {author} {\bibfnamefont {E.~S.}\ \bibnamefont {Toberer}},\ }\bibfield  {title} {\bibinfo {title} {New kagome prototype materials: discovery of {KV}$_{3}${Sb}$_{5}$,{RbV}$_{3}${Sb}$_{5}$ ,
  and {CsV}$_{3}${Sb}$_{5}$},\ }\href {https://doi.org/10.1103/physrevmaterials.3.094407} {\bibfield  {journal} {\bibinfo  {journal} {Physical Review Materials}\ }\textbf {\bibinfo {volume} {3}},\ \bibinfo {pages} {094407} (\bibinfo {year} {2019})}\BibitemShut {NoStop}%
\bibitem [{\citenamefont {Wilson}\ and\ \citenamefont {Ortiz}(2024)}]{Wilson2024_AV3Sb5-Review}%
  \BibitemOpen
  \bibfield  {author} {\bibinfo {author} {\bibfnamefont {S.~D.}\ \bibnamefont {Wilson}}\ and\ \bibinfo {author} {\bibfnamefont {B.~R.}\ \bibnamefont {Ortiz}},\ }\bibfield  {title} {\bibinfo {title} {${AV}_{3}${Sb}$_{5}$ kagome superconductors},\ }\href {https://doi.org/10.1038/s41578-024-00677-y} {\bibfield  {journal} {\bibinfo  {journal} {Nat. Rev. Mater.}\ ,\ \bibinfo {pages} {9756}} (\bibinfo {year} {2024})}\BibitemShut {NoStop}%
\bibitem [{\citenamefont {Li}\ \emph {et~al.}(2025)\citenamefont {Li}, \citenamefont {Liu}, \citenamefont {Du}, \citenamefont {Wu}, \citenamefont {Zhao}, \citenamefont {Zhai}, \citenamefont {Hu}, \citenamefont {Zhang}, \citenamefont {Chen}, \citenamefont {Liu}, \citenamefont {Yang}, \citenamefont {Peng}, \citenamefont {Hashimoto}, \citenamefont {Lu}, \citenamefont {Liu}, \citenamefont {Wang}, \citenamefont {Chen}, \citenamefont {Cao},\ and\ \citenamefont {Yang}}]{Li2025_CsCr3Sb5-ElectronicStructure+FlatBands}%
  \BibitemOpen
  \bibfield  {author} {\bibinfo {author} {\bibfnamefont {Y.}~\bibnamefont {Li}}, \bibinfo {author} {\bibfnamefont {Y.}~\bibnamefont {Liu}}, \bibinfo {author} {\bibfnamefont {X.}~\bibnamefont {Du}}, \bibinfo {author} {\bibfnamefont {S.}~\bibnamefont {Wu}}, \bibinfo {author} {\bibfnamefont {W.}~\bibnamefont {Zhao}}, \bibinfo {author} {\bibfnamefont {K.}~\bibnamefont {Zhai}}, \bibinfo {author} {\bibfnamefont {Y.}~\bibnamefont {Hu}}, \bibinfo {author} {\bibfnamefont {S.}~\bibnamefont {Zhang}}, \bibinfo {author} {\bibfnamefont {H.}~\bibnamefont {Chen}}, \bibinfo {author} {\bibfnamefont {J.}~\bibnamefont {Liu}}, \bibinfo {author} {\bibfnamefont {Y.}~\bibnamefont {Yang}}, \bibinfo {author} {\bibfnamefont {C.}~\bibnamefont {Peng}}, \bibinfo {author} {\bibfnamefont {M.}~\bibnamefont {Hashimoto}}, \bibinfo {author} {\bibfnamefont {D.}~\bibnamefont {Lu}}, \bibinfo {author} {\bibfnamefont {Z.}~\bibnamefont {Liu}}, \bibinfo {author} {\bibfnamefont {Y.}~\bibnamefont {Wang}}, \bibinfo {author} {\bibfnamefont
  {Y.}~\bibnamefont {Chen}}, \bibinfo {author} {\bibfnamefont {G.}~\bibnamefont {Cao}},\ and\ \bibinfo {author} {\bibfnamefont {L.}~\bibnamefont {Yang}},\ }\bibfield  {title} {\bibinfo {title} {Electron correlation and incipient flat bands in the {K}agome superconductor {CsCr}$_{3}${Sb}$_{5}$},\ }\bibfield  {journal} {\bibinfo  {journal} {Nature Communications}\ }\textbf {\bibinfo {volume} {16}},\ \href {https://doi.org/10.1038/s41467-025-58487-x} {10.1038/s41467-025-58487-x} (\bibinfo {year} {2025})\BibitemShut {NoStop}%
\bibitem [{\citenamefont {Wang}\ \emph {et~al.}(2022{\natexlab{a}})\citenamefont {Wang}, \citenamefont {Gu}, \citenamefont {McGuire}, \citenamefont {Yan}, \citenamefont {Shi}, \citenamefont {Cui}, \citenamefont {Chen}, \citenamefont {Wang}, \citenamefont {Zhang}, \citenamefont {Yang}, \citenamefont {Dong}, \citenamefont {Jiang}, \citenamefont {Hu}, \citenamefont {Wang}, \citenamefont {Sun},\ and\ \citenamefont {Cheng}}]{Wang2022_DensityWave-V3Sb2}%
  \BibitemOpen
  \bibfield  {author} {\bibinfo {author} {\bibfnamefont {N.}~\bibnamefont {Wang}}, \bibinfo {author} {\bibfnamefont {Y.}~\bibnamefont {Gu}}, \bibinfo {author} {\bibfnamefont {M.~A.}\ \bibnamefont {McGuire}}, \bibinfo {author} {\bibfnamefont {J.}~\bibnamefont {Yan}}, \bibinfo {author} {\bibfnamefont {L.}~\bibnamefont {Shi}}, \bibinfo {author} {\bibfnamefont {Q.}~\bibnamefont {Cui}}, \bibinfo {author} {\bibfnamefont {K.}~\bibnamefont {Chen}}, \bibinfo {author} {\bibfnamefont {Y.}~\bibnamefont {Wang}}, \bibinfo {author} {\bibfnamefont {H.}~\bibnamefont {Zhang}}, \bibinfo {author} {\bibfnamefont {H.}~\bibnamefont {Yang}}, \bibinfo {author} {\bibfnamefont {X.}~\bibnamefont {Dong}}, \bibinfo {author} {\bibfnamefont {K.}~\bibnamefont {Jiang}}, \bibinfo {author} {\bibfnamefont {J.}~\bibnamefont {Hu}}, \bibinfo {author} {\bibfnamefont {B.}~\bibnamefont {Wang}}, \bibinfo {author} {\bibfnamefont {J.}~\bibnamefont {Sun}},\ and\ \bibinfo {author} {\bibfnamefont {J.}~\bibnamefont {Cheng}},\ }\bibfield  {title} {\bibinfo
  {title} {A density-wave-like transition in the polycrystalline {V}$_{3}${S}b$_{2}$ sample with bilayer kagome lattice},\ }\href {https://doi.org/10.1088/1674-1056/ac4227} {\bibfield  {journal} {\bibinfo  {journal} {Chinese Physics B}\ }\textbf {\bibinfo {volume} {31}},\ \bibinfo {pages} {017106} (\bibinfo {year} {2022}{\natexlab{a}})}\BibitemShut {NoStop}%
\bibitem [{\citenamefont {Plokhikh}\ \emph {et~al.}(2024)\citenamefont {Plokhikh}, \citenamefont {Mielke}, \citenamefont {Nakamura}, \citenamefont {Petricek}, \citenamefont {Qin}, \citenamefont {Sazgari}, \citenamefont {Küspert}, \citenamefont {Biało}, \citenamefont {Shin}, \citenamefont {Ivashko}, \citenamefont {Graham}, \citenamefont {Zimmermann}, \citenamefont {Medarde}, \citenamefont {Amato}, \citenamefont {Khasanov}, \citenamefont {Luetkens}, \citenamefont {Fischer}, \citenamefont {Hasan}, \citenamefont {Yin}, \citenamefont {Neupert}, \citenamefont {Chang}, \citenamefont {Xu}, \citenamefont {Nakatsuji}, \citenamefont {Pomjakushina}, \citenamefont {Gawryluk},\ and\ \citenamefont {Guguchia}}]{Plokhikh2024_CDW-LuRu1-xFex3Si2}%
  \BibitemOpen
  \bibfield  {author} {\bibinfo {author} {\bibfnamefont {I.}~\bibnamefont {Plokhikh}}, \bibinfo {author} {\bibfnamefont {C.}~\bibnamefont {Mielke}}, \bibinfo {author} {\bibfnamefont {H.}~\bibnamefont {Nakamura}}, \bibinfo {author} {\bibfnamefont {V.}~\bibnamefont {Petricek}}, \bibinfo {author} {\bibfnamefont {Y.}~\bibnamefont {Qin}}, \bibinfo {author} {\bibfnamefont {V.}~\bibnamefont {Sazgari}}, \bibinfo {author} {\bibfnamefont {J.}~\bibnamefont {Küspert}}, \bibinfo {author} {\bibfnamefont {I.}~\bibnamefont {Biało}}, \bibinfo {author} {\bibfnamefont {S.}~\bibnamefont {Shin}}, \bibinfo {author} {\bibfnamefont {O.}~\bibnamefont {Ivashko}}, \bibinfo {author} {\bibfnamefont {J.~N.}\ \bibnamefont {Graham}}, \bibinfo {author} {\bibfnamefont {M.~v.}\ \bibnamefont {Zimmermann}}, \bibinfo {author} {\bibfnamefont {M.}~\bibnamefont {Medarde}}, \bibinfo {author} {\bibfnamefont {A.}~\bibnamefont {Amato}}, \bibinfo {author} {\bibfnamefont {R.}~\bibnamefont {Khasanov}}, \bibinfo {author} {\bibfnamefont {H.}~\bibnamefont
  {Luetkens}}, \bibinfo {author} {\bibfnamefont {M.~H.}\ \bibnamefont {Fischer}}, \bibinfo {author} {\bibfnamefont {M.~Z.}\ \bibnamefont {Hasan}}, \bibinfo {author} {\bibfnamefont {J.-X.}\ \bibnamefont {Yin}}, \bibinfo {author} {\bibfnamefont {T.}~\bibnamefont {Neupert}}, \bibinfo {author} {\bibfnamefont {J.}~\bibnamefont {Chang}}, \bibinfo {author} {\bibfnamefont {G.}~\bibnamefont {Xu}}, \bibinfo {author} {\bibfnamefont {S.}~\bibnamefont {Nakatsuji}}, \bibinfo {author} {\bibfnamefont {E.}~\bibnamefont {Pomjakushina}}, \bibinfo {author} {\bibfnamefont {D.~J.}\ \bibnamefont {Gawryluk}},\ and\ \bibinfo {author} {\bibfnamefont {Z.}~\bibnamefont {Guguchia}},\ }\bibfield  {title} {\bibinfo {title} {Discovery of charge order above room-temperature in the prototypical kagome superconductor {La(Ru}$_{1-x}${Fe}$_{x}$)$_{3}${Si}$_{2}$},\ }\bibfield  {journal} {\bibinfo  {journal} {Communications Physics}\ }\textbf {\bibinfo {volume} {7}},\ \href {https://doi.org/10.1038/s42005-024-01673-y} {10.1038/s42005-024-01673-y}
  (\bibinfo {year} {2024})\BibitemShut {NoStop}%
\bibitem [{\citenamefont {Wang}\ \emph {et~al.}(2022{\natexlab{b}})\citenamefont {Wang}, \citenamefont {McCandless}, \citenamefont {Wang}, \citenamefont {Thanabalasingam}, \citenamefont {Wu}, \citenamefont {Bouwmeester}, \citenamefont {van~der Zant}, \citenamefont {Ali},\ and\ \citenamefont {Chan}}]{Wang2022_Properties+PhaseTransitionYb0.5Co3Ge3}%
  \BibitemOpen
  \bibfield  {author} {\bibinfo {author} {\bibfnamefont {Y.}~\bibnamefont {Wang}}, \bibinfo {author} {\bibfnamefont {G.~T.}\ \bibnamefont {McCandless}}, \bibinfo {author} {\bibfnamefont {X.}~\bibnamefont {Wang}}, \bibinfo {author} {\bibfnamefont {K.}~\bibnamefont {Thanabalasingam}}, \bibinfo {author} {\bibfnamefont {H.}~\bibnamefont {Wu}}, \bibinfo {author} {\bibfnamefont {D.}~\bibnamefont {Bouwmeester}}, \bibinfo {author} {\bibfnamefont {H.~S.~J.}\ \bibnamefont {van~der Zant}}, \bibinfo {author} {\bibfnamefont {M.~N.}\ \bibnamefont {Ali}},\ and\ \bibinfo {author} {\bibfnamefont {J.~Y.}\ \bibnamefont {Chan}},\ }\bibfield  {title} {\bibinfo {title} {Electronic properties and phase transition in the kagome metal {Yb}$_{0.5}${Co}$_{3}${Ge}$_{3}$},\ }\href {https://doi.org/10.1021/acs.chemmater.2c01309} {\bibfield  {journal} {\bibinfo  {journal} {Chemistry of Materials}\ }\textbf {\bibinfo {volume} {34}},\ \bibinfo {pages} {7337} (\bibinfo {year} {2022}{\natexlab{b}})}\BibitemShut {NoStop}%
\bibitem [{\citenamefont {Teng}\ \emph {et~al.}(2023)\citenamefont {Teng}, \citenamefont {Oh}, \citenamefont {Tan}, \citenamefont {Chen}, \citenamefont {Huang}, \citenamefont {Gao}, \citenamefont {Yin}, \citenamefont {Chu}, \citenamefont {Hashimoto}, \citenamefont {Lu}, \citenamefont {Jozwiak}, \citenamefont {Bostwick}, \citenamefont {Rotenberg}, \citenamefont {Granroth}, \citenamefont {Yan}, \citenamefont {Birgeneau}, \citenamefont {Dai},\ and\ \citenamefont {Yi}}]{Teng2023_Mag+CDW-inHexFeGe}%
  \BibitemOpen
  \bibfield  {author} {\bibinfo {author} {\bibfnamefont {X.}~\bibnamefont {Teng}}, \bibinfo {author} {\bibfnamefont {J.~S.}\ \bibnamefont {Oh}}, \bibinfo {author} {\bibfnamefont {H.}~\bibnamefont {Tan}}, \bibinfo {author} {\bibfnamefont {L.}~\bibnamefont {Chen}}, \bibinfo {author} {\bibfnamefont {J.}~\bibnamefont {Huang}}, \bibinfo {author} {\bibfnamefont {B.}~\bibnamefont {Gao}}, \bibinfo {author} {\bibfnamefont {J.-X.}\ \bibnamefont {Yin}}, \bibinfo {author} {\bibfnamefont {J.-H.}\ \bibnamefont {Chu}}, \bibinfo {author} {\bibfnamefont {M.}~\bibnamefont {Hashimoto}}, \bibinfo {author} {\bibfnamefont {D.}~\bibnamefont {Lu}}, \bibinfo {author} {\bibfnamefont {C.}~\bibnamefont {Jozwiak}}, \bibinfo {author} {\bibfnamefont {A.}~\bibnamefont {Bostwick}}, \bibinfo {author} {\bibfnamefont {E.}~\bibnamefont {Rotenberg}}, \bibinfo {author} {\bibfnamefont {G.~E.}\ \bibnamefont {Granroth}}, \bibinfo {author} {\bibfnamefont {B.}~\bibnamefont {Yan}}, \bibinfo {author} {\bibfnamefont {R.~J.}\ \bibnamefont {Birgeneau}},
  \bibinfo {author} {\bibfnamefont {P.}~\bibnamefont {Dai}},\ and\ \bibinfo {author} {\bibfnamefont {M.}~\bibnamefont {Yi}},\ }\bibfield  {title} {\bibinfo {title} {Magnetism and charge density wave order in kagome {FeGe}},\ }\href {https://doi.org/10.1038/s41567-023-01985-w} {\bibfield  {journal} {\bibinfo  {journal} {Nature Physics}\ }\textbf {\bibinfo {volume} {19}},\ \bibinfo {pages} {814} (\bibinfo {year} {2023})}\BibitemShut {NoStop}%
\bibitem [{\citenamefont {Teng}\ \emph {et~al.}(2022)\citenamefont {Teng}, \citenamefont {Chen}, \citenamefont {Ye}, \citenamefont {Rosenberg}, \citenamefont {Liu}, \citenamefont {Yin}, \citenamefont {Jiang}, \citenamefont {Oh}, \citenamefont {Hasan}, \citenamefont {Neubauer}, \citenamefont {Gao}, \citenamefont {Xie}, \citenamefont {Hashimoto}, \citenamefont {Lu}, \citenamefont {Jozwiak}, \citenamefont {Bostwick}, \citenamefont {Rotenberg}, \citenamefont {Birgeneau}, \citenamefont {Chu}, \citenamefont {Yi},\ and\ \citenamefont {Dai}}]{Teng2022_CDWinHexFeGe}%
  \BibitemOpen
  \bibfield  {author} {\bibinfo {author} {\bibfnamefont {X.}~\bibnamefont {Teng}}, \bibinfo {author} {\bibfnamefont {L.}~\bibnamefont {Chen}}, \bibinfo {author} {\bibfnamefont {F.}~\bibnamefont {Ye}}, \bibinfo {author} {\bibfnamefont {E.}~\bibnamefont {Rosenberg}}, \bibinfo {author} {\bibfnamefont {Z.}~\bibnamefont {Liu}}, \bibinfo {author} {\bibfnamefont {J.-X.}\ \bibnamefont {Yin}}, \bibinfo {author} {\bibfnamefont {Y.-X.}\ \bibnamefont {Jiang}}, \bibinfo {author} {\bibfnamefont {J.~S.}\ \bibnamefont {Oh}}, \bibinfo {author} {\bibfnamefont {M.~Z.}\ \bibnamefont {Hasan}}, \bibinfo {author} {\bibfnamefont {K.~J.}\ \bibnamefont {Neubauer}}, \bibinfo {author} {\bibfnamefont {B.}~\bibnamefont {Gao}}, \bibinfo {author} {\bibfnamefont {Y.}~\bibnamefont {Xie}}, \bibinfo {author} {\bibfnamefont {M.}~\bibnamefont {Hashimoto}}, \bibinfo {author} {\bibfnamefont {D.}~\bibnamefont {Lu}}, \bibinfo {author} {\bibfnamefont {C.}~\bibnamefont {Jozwiak}}, \bibinfo {author} {\bibfnamefont {A.}~\bibnamefont {Bostwick}}, \bibinfo
  {author} {\bibfnamefont {E.}~\bibnamefont {Rotenberg}}, \bibinfo {author} {\bibfnamefont {R.~J.}\ \bibnamefont {Birgeneau}}, \bibinfo {author} {\bibfnamefont {J.-H.}\ \bibnamefont {Chu}}, \bibinfo {author} {\bibfnamefont {M.}~\bibnamefont {Yi}},\ and\ \bibinfo {author} {\bibfnamefont {P.}~\bibnamefont {Dai}},\ }\bibfield  {title} {\bibinfo {title} {Discovery of charge density wave in a kagome lattice antiferromagnet},\ }\href {https://doi.org/10.1038/s41586-022-05034-z} {\bibfield  {journal} {\bibinfo  {journal} {Nature}\ }\textbf {\bibinfo {volume} {609}},\ \bibinfo {pages} {490} (\bibinfo {year} {2022})}\BibitemShut {NoStop}%
\bibitem [{\citenamefont {Chen}\ \emph {et~al.}(2024)\citenamefont {Chen}, \citenamefont {Wu}, \citenamefont {Zhou}, \citenamefont {Zhang}, \citenamefont {Yin}, \citenamefont {Li}, \citenamefont {Li}, \citenamefont {Gong}, \citenamefont {He}, \citenamefont {Chai}, \citenamefont {Zhou}, \citenamefont {Wang}, \citenamefont {Wang}, \citenamefont {Yan},\ and\ \citenamefont {Feng}}]{Chen2024_LongRangeCDW+DimerizationHexFeGe}%
  \BibitemOpen
  \bibfield  {author} {\bibinfo {author} {\bibfnamefont {Z.}~\bibnamefont {Chen}}, \bibinfo {author} {\bibfnamefont {X.}~\bibnamefont {Wu}}, \bibinfo {author} {\bibfnamefont {S.}~\bibnamefont {Zhou}}, \bibinfo {author} {\bibfnamefont {J.}~\bibnamefont {Zhang}}, \bibinfo {author} {\bibfnamefont {R.}~\bibnamefont {Yin}}, \bibinfo {author} {\bibfnamefont {Y.}~\bibnamefont {Li}}, \bibinfo {author} {\bibfnamefont {M.}~\bibnamefont {Li}}, \bibinfo {author} {\bibfnamefont {J.}~\bibnamefont {Gong}}, \bibinfo {author} {\bibfnamefont {M.}~\bibnamefont {He}}, \bibinfo {author} {\bibfnamefont {Y.}~\bibnamefont {Chai}}, \bibinfo {author} {\bibfnamefont {X.}~\bibnamefont {Zhou}}, \bibinfo {author} {\bibfnamefont {Y.}~\bibnamefont {Wang}}, \bibinfo {author} {\bibfnamefont {A.}~\bibnamefont {Wang}}, \bibinfo {author} {\bibfnamefont {Y.-J.}\ \bibnamefont {Yan}},\ and\ \bibinfo {author} {\bibfnamefont {D.-L.}\ \bibnamefont {Feng}},\ }\bibfield  {title} {\bibinfo {title} {Discovery of a long-ranged charge order with 1/4
  {Ge}1-dimerization in an antiferromagnetic kagome metal},\ }\bibfield  {journal} {\bibinfo  {journal} {Nature Communications}\ }\textbf {\bibinfo {volume} {15}},\ \href {https://doi.org/10.1038/s41467-024-50661-x} {10.1038/s41467-024-50661-x} (\bibinfo {year} {2024})\BibitemShut {NoStop}%
\bibitem [{\citenamefont {Arachchige}\ \emph {et~al.}(2022)\citenamefont {Arachchige}, \citenamefont {Meier}, \citenamefont {Marshall}, \citenamefont {Matsuoka}, \citenamefont {Xue}, \citenamefont {McGuire}, \citenamefont {Hermann}, \citenamefont {Cao},\ and\ \citenamefont {Mandrus}}]{Arachchige2022_CDW-ScV6Sn6}%
  \BibitemOpen
  \bibfield  {author} {\bibinfo {author} {\bibfnamefont {H.~W.~S.}\ \bibnamefont {Arachchige}}, \bibinfo {author} {\bibfnamefont {W.~R.}\ \bibnamefont {Meier}}, \bibinfo {author} {\bibfnamefont {M.}~\bibnamefont {Marshall}}, \bibinfo {author} {\bibfnamefont {T.}~\bibnamefont {Matsuoka}}, \bibinfo {author} {\bibfnamefont {R.}~\bibnamefont {Xue}}, \bibinfo {author} {\bibfnamefont {M.~A.}\ \bibnamefont {McGuire}}, \bibinfo {author} {\bibfnamefont {R.~P.}\ \bibnamefont {Hermann}}, \bibinfo {author} {\bibfnamefont {H.}~\bibnamefont {Cao}},\ and\ \bibinfo {author} {\bibfnamefont {D.}~\bibnamefont {Mandrus}},\ }\bibfield  {title} {\bibinfo {title} {Charge density wave in kagome lattice intermetallic {ScV}$_{6}${Sn}$_{6}$},\ }\href {https://doi.org/10.1103/PhysRevLett.129.216402} {\bibfield  {journal} {\bibinfo  {journal} {Physical Review Letters}\ }\textbf {\bibinfo {volume} {129}},\ \bibinfo {pages} {216402} (\bibinfo {year} {2022})}\BibitemShut {NoStop}%
\bibitem [{\citenamefont {Mozaffari}\ \emph {et~al.}(2024)\citenamefont {Mozaffari}, \citenamefont {Meier}, \citenamefont {Madhogaria}, \citenamefont {Peshcherenko}, \citenamefont {Kang}, \citenamefont {Villanova}, \citenamefont {Arachchige}, \citenamefont {Zheng}, \citenamefont {Zhu}, \citenamefont {Chen}, \citenamefont {Jenkins}, \citenamefont {Zhang}, \citenamefont {Chan}, \citenamefont {Li}, \citenamefont {Yoon}, \citenamefont {Zhang},\ and\ \citenamefont {Mandrus}}]{Mozaffari2024_RV6Sn6-SublinearResistivity}%
  \BibitemOpen
  \bibfield  {author} {\bibinfo {author} {\bibfnamefont {S.}~\bibnamefont {Mozaffari}}, \bibinfo {author} {\bibfnamefont {W.~R.}\ \bibnamefont {Meier}}, \bibinfo {author} {\bibfnamefont {R.~P.}\ \bibnamefont {Madhogaria}}, \bibinfo {author} {\bibfnamefont {N.}~\bibnamefont {Peshcherenko}}, \bibinfo {author} {\bibfnamefont {S.-H.}\ \bibnamefont {Kang}}, \bibinfo {author} {\bibfnamefont {J.~W.}\ \bibnamefont {Villanova}}, \bibinfo {author} {\bibfnamefont {H.~W.~S.}\ \bibnamefont {Arachchige}}, \bibinfo {author} {\bibfnamefont {G.}~\bibnamefont {Zheng}}, \bibinfo {author} {\bibfnamefont {Y.}~\bibnamefont {Zhu}}, \bibinfo {author} {\bibfnamefont {K.-W.}\ \bibnamefont {Chen}}, \bibinfo {author} {\bibfnamefont {K.}~\bibnamefont {Jenkins}}, \bibinfo {author} {\bibfnamefont {D.}~\bibnamefont {Zhang}}, \bibinfo {author} {\bibfnamefont {A.}~\bibnamefont {Chan}}, \bibinfo {author} {\bibfnamefont {L.}~\bibnamefont {Li}}, \bibinfo {author} {\bibfnamefont {M.}~\bibnamefont {Yoon}}, \bibinfo {author} {\bibfnamefont
  {Y.}~\bibnamefont {Zhang}},\ and\ \bibinfo {author} {\bibfnamefont {D.~G.}\ \bibnamefont {Mandrus}},\ }\bibfield  {title} {\bibinfo {title} {Universal sublinear resistivity in vanadium kagome materials hosting charge density waves},\ }\href {https://doi.org/10.1103/physrevb.110.035135} {\bibfield  {journal} {\bibinfo  {journal} {Physical Review B}\ }\textbf {\bibinfo {volume} {110}},\ \bibinfo {pages} {035135} (\bibinfo {year} {2024})}\BibitemShut {NoStop}%
\bibitem [{\citenamefont {DeStefano}\ \emph {et~al.}(2023)\citenamefont {DeStefano}, \citenamefont {Rosenberg}, \citenamefont {Peek}, \citenamefont {Lee}, \citenamefont {Liu}, \citenamefont {Jiang}, \citenamefont {Ke},\ and\ \citenamefont {Chu}}]{DeStefano2023_ScV6Sn6-TransportRevealedByMagTrasport}%
  \BibitemOpen
  \bibfield  {author} {\bibinfo {author} {\bibfnamefont {J.~M.}\ \bibnamefont {DeStefano}}, \bibinfo {author} {\bibfnamefont {E.}~\bibnamefont {Rosenberg}}, \bibinfo {author} {\bibfnamefont {O.}~\bibnamefont {Peek}}, \bibinfo {author} {\bibfnamefont {Y.}~\bibnamefont {Lee}}, \bibinfo {author} {\bibfnamefont {Z.}~\bibnamefont {Liu}}, \bibinfo {author} {\bibfnamefont {Q.}~\bibnamefont {Jiang}}, \bibinfo {author} {\bibfnamefont {L.}~\bibnamefont {Ke}},\ and\ \bibinfo {author} {\bibfnamefont {J.-H.}\ \bibnamefont {Chu}},\ }\bibfield  {title} {\bibinfo {title} {Pseudogap behavior in charge density wave kagome material {ScV}$_{6}${Sn}$_{6}$ revealed by magnetotransport measurements},\ }\bibfield  {journal} {\bibinfo  {journal} {npj Quantum Materials}\ }\textbf {\bibinfo {volume} {8}},\ \href {https://doi.org/10.1038/s41535-023-00600-8} {10.1038/s41535-023-00600-8} (\bibinfo {year} {2023})\BibitemShut {NoStop}%
\bibitem [{\citenamefont {Yi}\ \emph {et~al.}(2024)\citenamefont {Yi}, \citenamefont {Feng}, \citenamefont {Mao}, \citenamefont {Yanda}, \citenamefont {Roychowdhury}, \citenamefont {Zhang}, \citenamefont {Felser},\ and\ \citenamefont {Shekhar}}]{Yi2024_QuantumOscillationsScV6Sn6}%
  \BibitemOpen
  \bibfield  {author} {\bibinfo {author} {\bibfnamefont {C.}~\bibnamefont {Yi}}, \bibinfo {author} {\bibfnamefont {X.}~\bibnamefont {Feng}}, \bibinfo {author} {\bibfnamefont {N.}~\bibnamefont {Mao}}, \bibinfo {author} {\bibfnamefont {P.}~\bibnamefont {Yanda}}, \bibinfo {author} {\bibfnamefont {S.}~\bibnamefont {Roychowdhury}}, \bibinfo {author} {\bibfnamefont {Y.}~\bibnamefont {Zhang}}, \bibinfo {author} {\bibfnamefont {C.}~\bibnamefont {Felser}},\ and\ \bibinfo {author} {\bibfnamefont {C.}~\bibnamefont {Shekhar}},\ }\bibfield  {title} {\bibinfo {title} {Quantum oscillations revealing topological band in kagome metal {ScV}$_{6}${Sn}$_{6}$},\ }\href {https://doi.org/10.1103/physrevb.109.035124} {\bibfield  {journal} {\bibinfo  {journal} {Physical Review B}\ }\textbf {\bibinfo {volume} {109}},\ \bibinfo {pages} {035124} (\bibinfo {year} {2024})}\BibitemShut {NoStop}%
\bibitem [{\citenamefont {Kuo}\ \emph {et~al.}(2024)\citenamefont {Kuo}, \citenamefont {Huang}, \citenamefont {Tian}, \citenamefont {Hong}, \citenamefont {Ou}, \citenamefont {Kuo},\ and\ \citenamefont {Lue}}]{Kuo2024_ScV6Sn6-Termoelectric}%
  \BibitemOpen
  \bibfield  {author} {\bibinfo {author} {\bibfnamefont {C.~N.}\ \bibnamefont {Kuo}}, \bibinfo {author} {\bibfnamefont {R.~Y.}\ \bibnamefont {Huang}}, \bibinfo {author} {\bibfnamefont {W.~S.}\ \bibnamefont {Tian}}, \bibinfo {author} {\bibfnamefont {C.~K.}\ \bibnamefont {Hong}}, \bibinfo {author} {\bibfnamefont {Y.~R.}\ \bibnamefont {Ou}}, \bibinfo {author} {\bibfnamefont {Y.~K.}\ \bibnamefont {Kuo}},\ and\ \bibinfo {author} {\bibfnamefont {C.~S.}\ \bibnamefont {Lue}},\ }\bibfield  {title} {\bibinfo {title} {Effects of lattice instability on the thermoelectric behavior of kagome metal {ScV}$_{6}${Sn}$_{6}$},\ }\bibfield  {journal} {\bibinfo  {journal} {Applied Physics Letters}\ }\textbf {\bibinfo {volume} {125}},\ \href {https://doi.org/10.1063/5.0231041} {10.1063/5.0231041} (\bibinfo {year} {2024})\BibitemShut {NoStop}%
\bibitem [{\citenamefont {Cao}\ \emph {et~al.}(2023)\citenamefont {Cao}, \citenamefont {Xu}, \citenamefont {Fukui}, \citenamefont {Manjo}, \citenamefont {Dong}, \citenamefont {Shi}, \citenamefont {Liu}, \citenamefont {Cao},\ and\ \citenamefont {Song}}]{Cao2023_CompetingCDWs-ScV6Sn6}%
  \BibitemOpen
  \bibfield  {author} {\bibinfo {author} {\bibfnamefont {S.}~\bibnamefont {Cao}}, \bibinfo {author} {\bibfnamefont {C.}~\bibnamefont {Xu}}, \bibinfo {author} {\bibfnamefont {H.}~\bibnamefont {Fukui}}, \bibinfo {author} {\bibfnamefont {T.}~\bibnamefont {Manjo}}, \bibinfo {author} {\bibfnamefont {Y.}~\bibnamefont {Dong}}, \bibinfo {author} {\bibfnamefont {M.}~\bibnamefont {Shi}}, \bibinfo {author} {\bibfnamefont {Y.}~\bibnamefont {Liu}}, \bibinfo {author} {\bibfnamefont {C.}~\bibnamefont {Cao}},\ and\ \bibinfo {author} {\bibfnamefont {Y.}~\bibnamefont {Song}},\ }\bibfield  {title} {\bibinfo {title} {Competing charge-density wave instabilities in the kagome metal {ScV}$_{6}${Sn}$_{6}$},\ }\bibfield  {journal} {\bibinfo  {journal} {Nature Communications}\ }\textbf {\bibinfo {volume} {14}},\ \href {https://doi.org/10.1038/s41467-023-43454-1} {10.1038/s41467-023-43454-1} (\bibinfo {year} {2023})\BibitemShut {NoStop}%
\bibitem [{\citenamefont {Korshunov}\ \emph {et~al.}(2023)\citenamefont {Korshunov}, \citenamefont {Hu}, \citenamefont {Subires}, \citenamefont {Jiang}, \citenamefont {Călugăru}, \citenamefont {Feng}, \citenamefont {Rajapitamahuni}, \citenamefont {Yi}, \citenamefont {Roychowdhury}, \citenamefont {Vergniory}, \citenamefont {Strempfer}, \citenamefont {Shekhar}, \citenamefont {Vescovo}, \citenamefont {Chernyshov}, \citenamefont {Said}, \citenamefont {Bosak}, \citenamefont {Felser}, \citenamefont {Bernevig},\ and\ \citenamefont {Blanco-Canosa}}]{Korshunov2023_SofteningFlatPhononScV6Sn6}%
  \BibitemOpen
  \bibfield  {author} {\bibinfo {author} {\bibfnamefont {A.}~\bibnamefont {Korshunov}}, \bibinfo {author} {\bibfnamefont {H.}~\bibnamefont {Hu}}, \bibinfo {author} {\bibfnamefont {D.}~\bibnamefont {Subires}}, \bibinfo {author} {\bibfnamefont {Y.}~\bibnamefont {Jiang}}, \bibinfo {author} {\bibfnamefont {D.}~\bibnamefont {Călugăru}}, \bibinfo {author} {\bibfnamefont {X.}~\bibnamefont {Feng}}, \bibinfo {author} {\bibfnamefont {A.}~\bibnamefont {Rajapitamahuni}}, \bibinfo {author} {\bibfnamefont {C.}~\bibnamefont {Yi}}, \bibinfo {author} {\bibfnamefont {S.}~\bibnamefont {Roychowdhury}}, \bibinfo {author} {\bibfnamefont {M.~G.}\ \bibnamefont {Vergniory}}, \bibinfo {author} {\bibfnamefont {J.}~\bibnamefont {Strempfer}}, \bibinfo {author} {\bibfnamefont {C.}~\bibnamefont {Shekhar}}, \bibinfo {author} {\bibfnamefont {E.}~\bibnamefont {Vescovo}}, \bibinfo {author} {\bibfnamefont {D.}~\bibnamefont {Chernyshov}}, \bibinfo {author} {\bibfnamefont {A.~H.}\ \bibnamefont {Said}}, \bibinfo {author} {\bibfnamefont
  {A.}~\bibnamefont {Bosak}}, \bibinfo {author} {\bibfnamefont {C.}~\bibnamefont {Felser}}, \bibinfo {author} {\bibfnamefont {B.~A.}\ \bibnamefont {Bernevig}},\ and\ \bibinfo {author} {\bibfnamefont {S.}~\bibnamefont {Blanco-Canosa}},\ }\bibfield  {title} {\bibinfo {title} {Softening of a flat phonon mode in the kagome {ScV}$_{6}${Sn}$_{6}$},\ }\bibfield  {journal} {\bibinfo  {journal} {Nature Communications}\ }\textbf {\bibinfo {volume} {14}},\ \href {https://doi.org/10.1038/s41467-023-42186-6} {10.1038/s41467-023-42186-6} (\bibinfo {year} {2023})\BibitemShut {NoStop}%
\bibitem [{\citenamefont {Hu}\ \emph {et~al.}(2025)\citenamefont {Hu}, \citenamefont {Jiang}, \citenamefont {C\u{a}lug\u{a}ru}, \citenamefont {Feng}, \citenamefont {Subires}, \citenamefont {Vergniory}, \citenamefont {Felser}, \citenamefont {Blanco-Canosa},\ and\ \citenamefont {Bernevig}}]{Hu2025_FlatPhononSoftModes-ScV6Sn6-PhononCalcs}%
  \BibitemOpen
  \bibfield  {author} {\bibinfo {author} {\bibfnamefont {H.}~\bibnamefont {Hu}}, \bibinfo {author} {\bibfnamefont {Y.}~\bibnamefont {Jiang}}, \bibinfo {author} {\bibfnamefont {D.}~\bibnamefont {C\u{a}lug\u{a}ru}}, \bibinfo {author} {\bibfnamefont {X.}~\bibnamefont {Feng}}, \bibinfo {author} {\bibfnamefont {D.}~\bibnamefont {Subires}}, \bibinfo {author} {\bibfnamefont {M.~G.}\ \bibnamefont {Vergniory}}, \bibinfo {author} {\bibfnamefont {C.}~\bibnamefont {Felser}}, \bibinfo {author} {\bibfnamefont {S.}~\bibnamefont {Blanco-Canosa}},\ and\ \bibinfo {author} {\bibfnamefont {B.~A.}\ \bibnamefont {Bernevig}},\ }\bibfield  {title} {\bibinfo {title} {Kagome materials {I}: {SG} 191, {ScV}$_{6}${Sn}$_{6}$. flat phonon soft modes and unconventional charge density wave formation: Microscopic and effective theory},\ }\href {https://doi.org/10.1103/physrevb.111.054113} {\bibfield  {journal} {\bibinfo  {journal} {Physical Review B}\ }\textbf {\bibinfo {volume} {111}},\ \bibinfo {pages} {054113} (\bibinfo {year}
  {2025})}\BibitemShut {NoStop}%
\bibitem [{\citenamefont {Pokharel}\ \emph {et~al.}(2023)\citenamefont {Pokharel}, \citenamefont {Ortiz}, \citenamefont {Kautzsch}, \citenamefont {Gomez~Alvarado}, \citenamefont {Mallayya}, \citenamefont {Wu}, \citenamefont {Kim}, \citenamefont {Ruff}, \citenamefont {Sarker},\ and\ \citenamefont {Wilson}}]{Pokharel2023_FrustratedChargeOrder+CooperativeDistortions-ScV6Sn6}%
  \BibitemOpen
  \bibfield  {author} {\bibinfo {author} {\bibfnamefont {G.}~\bibnamefont {Pokharel}}, \bibinfo {author} {\bibfnamefont {B.~R.}\ \bibnamefont {Ortiz}}, \bibinfo {author} {\bibfnamefont {L.}~\bibnamefont {Kautzsch}}, \bibinfo {author} {\bibfnamefont {S.~J.}\ \bibnamefont {Gomez~Alvarado}}, \bibinfo {author} {\bibfnamefont {K.}~\bibnamefont {Mallayya}}, \bibinfo {author} {\bibfnamefont {G.}~\bibnamefont {Wu}}, \bibinfo {author} {\bibfnamefont {E.-A.}\ \bibnamefont {Kim}}, \bibinfo {author} {\bibfnamefont {J.~P.~C.}\ \bibnamefont {Ruff}}, \bibinfo {author} {\bibfnamefont {S.}~\bibnamefont {Sarker}},\ and\ \bibinfo {author} {\bibfnamefont {S.~D.}\ \bibnamefont {Wilson}},\ }\bibfield  {title} {\bibinfo {title} {Frustrated charge order and cooperative distortions in {ScV}$_{6}${Sn}$_{6}$},\ }\href {https://doi.org/10.1103/physrevmaterials.7.104201} {\bibfield  {journal} {\bibinfo  {journal} {Physical Review Materials}\ }\textbf {\bibinfo {volume} {7}},\ \bibinfo {pages} {104201} (\bibinfo {year}
  {2023})}\BibitemShut {NoStop}%
\bibitem [{\citenamefont {Subedi}(2024)}]{Subedi2024_OrderByDisorderCDW-ScV6Sn6}%
  \BibitemOpen
  \bibfield  {author} {\bibinfo {author} {\bibfnamefont {A.}~\bibnamefont {Subedi}},\ }\bibfield  {title} {\bibinfo {title} {Order-by-disorder charge density wave condensation at $q$=(1/3,1/3,1/3) in kagome metal {ScV}$_{6}${Sn}$_{6}$},\ }\href {https://doi.org/10.1103/physrevmaterials.8.014006} {\bibfield  {journal} {\bibinfo  {journal} {Physical Review Materials}\ }\textbf {\bibinfo {volume} {8}},\ \bibinfo {pages} {014006} (\bibinfo {year} {2024})}\BibitemShut {NoStop}%
\bibitem [{\citenamefont {Wang}\ \emph {et~al.}(2024)\citenamefont {Wang}, \citenamefont {Chen}, \citenamefont {Kim},\ and\ \citenamefont {Monserrat}}]{Wang2024_OriginOfCompetingCDWs-ScV6Sn6}%
  \BibitemOpen
  \bibfield  {author} {\bibinfo {author} {\bibfnamefont {K.}~\bibnamefont {Wang}}, \bibinfo {author} {\bibfnamefont {S.}~\bibnamefont {Chen}}, \bibinfo {author} {\bibfnamefont {S.-W.}\ \bibnamefont {Kim}},\ and\ \bibinfo {author} {\bibfnamefont {B.}~\bibnamefont {Monserrat}},\ }\bibfield  {title} {\bibinfo {title} {Origin of competing charge density waves in kagome metal {ScV}$_{6}${Sn}$_{6}$},\ }\bibfield  {journal} {\bibinfo  {journal} {Nat Commun 15, 10428 (2024)}\ }\textbf {\bibinfo {volume} {15}},\ \href {https://doi.org/10.1038/s41467-024-54702-3} {10.1038/s41467-024-54702-3} (\bibinfo {year} {2024}),\ \Eprint {https://arxiv.org/abs/2403.17058} {2403.17058 [cond-mat.str-el]} \BibitemShut {NoStop}%
\bibitem [{\citenamefont {Tan}\ and\ \citenamefont {Yan}(2023)}]{Tan2023_AbundantLatticeInstabilites-ScV6Sn6}%
  \BibitemOpen
  \bibfield  {author} {\bibinfo {author} {\bibfnamefont {H.}~\bibnamefont {Tan}}\ and\ \bibinfo {author} {\bibfnamefont {B.}~\bibnamefont {Yan}},\ }\bibfield  {title} {\bibinfo {title} {Abundant lattice instability in kagome metal {ScV}$_{6}${Sn}$_{6}$},\ }\href {https://doi.org/10.1103/physrevlett.130.266402} {\bibfield  {journal} {\bibinfo  {journal} {Physical Review Letters}\ }\textbf {\bibinfo {volume} {130}},\ \bibinfo {pages} {266402} (\bibinfo {year} {2023})}\BibitemShut {NoStop}%
\bibitem [{\citenamefont {Tuniz}\ \emph {et~al.}(2023)\citenamefont {Tuniz}, \citenamefont {Consiglio}, \citenamefont {Puntel}, \citenamefont {Bigi}, \citenamefont {Enzner}, \citenamefont {Pokharel}, \citenamefont {Orgiani}, \citenamefont {Bronsch}, \citenamefont {Parmigiani}, \citenamefont {Polewczyk}, \citenamefont {King}, \citenamefont {Wells}, \citenamefont {Zeljkovic}, \citenamefont {Carrara}, \citenamefont {Rossi}, \citenamefont {Fujii}, \citenamefont {Vobornik}, \citenamefont {Wilson}, \citenamefont {Thomale}, \citenamefont {Wehling}, \citenamefont {Sangiovanni}, \citenamefont {Panaccione}, \citenamefont {Cilento}, \citenamefont {Di~Sante},\ and\ \citenamefont {Mazzola}}]{Tuniz2023_Dynamics+ResilienceOfCDW-ScV6Sn6}%
  \BibitemOpen
  \bibfield  {author} {\bibinfo {author} {\bibfnamefont {M.}~\bibnamefont {Tuniz}}, \bibinfo {author} {\bibfnamefont {A.}~\bibnamefont {Consiglio}}, \bibinfo {author} {\bibfnamefont {D.}~\bibnamefont {Puntel}}, \bibinfo {author} {\bibfnamefont {C.}~\bibnamefont {Bigi}}, \bibinfo {author} {\bibfnamefont {S.}~\bibnamefont {Enzner}}, \bibinfo {author} {\bibfnamefont {G.}~\bibnamefont {Pokharel}}, \bibinfo {author} {\bibfnamefont {P.}~\bibnamefont {Orgiani}}, \bibinfo {author} {\bibfnamefont {W.}~\bibnamefont {Bronsch}}, \bibinfo {author} {\bibfnamefont {F.}~\bibnamefont {Parmigiani}}, \bibinfo {author} {\bibfnamefont {V.}~\bibnamefont {Polewczyk}}, \bibinfo {author} {\bibfnamefont {P.~D.~C.}\ \bibnamefont {King}}, \bibinfo {author} {\bibfnamefont {J.~W.}\ \bibnamefont {Wells}}, \bibinfo {author} {\bibfnamefont {I.}~\bibnamefont {Zeljkovic}}, \bibinfo {author} {\bibfnamefont {P.}~\bibnamefont {Carrara}}, \bibinfo {author} {\bibfnamefont {G.}~\bibnamefont {Rossi}}, \bibinfo {author} {\bibfnamefont
  {J.}~\bibnamefont {Fujii}}, \bibinfo {author} {\bibfnamefont {I.}~\bibnamefont {Vobornik}}, \bibinfo {author} {\bibfnamefont {S.~D.}\ \bibnamefont {Wilson}}, \bibinfo {author} {\bibfnamefont {R.}~\bibnamefont {Thomale}}, \bibinfo {author} {\bibfnamefont {T.}~\bibnamefont {Wehling}}, \bibinfo {author} {\bibfnamefont {G.}~\bibnamefont {Sangiovanni}}, \bibinfo {author} {\bibfnamefont {G.}~\bibnamefont {Panaccione}}, \bibinfo {author} {\bibfnamefont {F.}~\bibnamefont {Cilento}}, \bibinfo {author} {\bibfnamefont {D.}~\bibnamefont {Di~Sante}},\ and\ \bibinfo {author} {\bibfnamefont {F.}~\bibnamefont {Mazzola}},\ }\bibfield  {title} {\bibinfo {title} {Dynamics and resilience of the unconventional charge density wave in {ScV}$_{6}${Sn}$_{6}$ bilayer kagome metal},\ }\bibfield  {journal} {\bibinfo  {journal} {Communications Materials}\ }\textbf {\bibinfo {volume} {4}},\ \href {https://doi.org/10.1038/s43246-023-00430-y} {10.1038/s43246-023-00430-y} (\bibinfo {year} {2023})\BibitemShut {NoStop}%
\bibitem [{\citenamefont {Lee}\ \emph {et~al.}(2024)\citenamefont {Lee}, \citenamefont {Won}, \citenamefont {Kim}, \citenamefont {Yoo}, \citenamefont {Park}, \citenamefont {Denlinger}, \citenamefont {Jozwiak}, \citenamefont {Bostwick}, \citenamefont {Rotenberg}, \citenamefont {Comin}, \citenamefont {Kang},\ and\ \citenamefont {Park}}]{Lee2024_NatureOfCDW-ScV6Sn6}%
  \BibitemOpen
  \bibfield  {author} {\bibinfo {author} {\bibfnamefont {S.}~\bibnamefont {Lee}}, \bibinfo {author} {\bibfnamefont {C.}~\bibnamefont {Won}}, \bibinfo {author} {\bibfnamefont {J.}~\bibnamefont {Kim}}, \bibinfo {author} {\bibfnamefont {J.}~\bibnamefont {Yoo}}, \bibinfo {author} {\bibfnamefont {S.}~\bibnamefont {Park}}, \bibinfo {author} {\bibfnamefont {J.}~\bibnamefont {Denlinger}}, \bibinfo {author} {\bibfnamefont {C.}~\bibnamefont {Jozwiak}}, \bibinfo {author} {\bibfnamefont {A.}~\bibnamefont {Bostwick}}, \bibinfo {author} {\bibfnamefont {E.}~\bibnamefont {Rotenberg}}, \bibinfo {author} {\bibfnamefont {R.}~\bibnamefont {Comin}}, \bibinfo {author} {\bibfnamefont {M.}~\bibnamefont {Kang}},\ and\ \bibinfo {author} {\bibfnamefont {J.-H.}\ \bibnamefont {Park}},\ }\bibfield  {title} {\bibinfo {title} {Nature of charge density wave in kagome metal {ScV}$_{6}${Sn}$_{6}$},\ }\bibfield  {journal} {\bibinfo  {journal} {npj Quantum Materials}\ }\textbf {\bibinfo {volume} {9}},\ \href
  {https://doi.org/10.1038/s41535-024-00620-y} {10.1038/s41535-024-00620-y} (\bibinfo {year} {2024})\BibitemShut {NoStop}%
\bibitem [{\citenamefont {Kim}\ \emph {et~al.}(2023)\citenamefont {Kim}, \citenamefont {Liu}, \citenamefont {Wang}, \citenamefont {Nam}, \citenamefont {Pokharel}, \citenamefont {Wilson}, \citenamefont {Cho},\ and\ \citenamefont {Moon}}]{Kim2023_IR-ProbeCDWGap-ScV6Sn6}%
  \BibitemOpen
  \bibfield  {author} {\bibinfo {author} {\bibfnamefont {D.~W.}\ \bibnamefont {Kim}}, \bibinfo {author} {\bibfnamefont {S.}~\bibnamefont {Liu}}, \bibinfo {author} {\bibfnamefont {C.}~\bibnamefont {Wang}}, \bibinfo {author} {\bibfnamefont {H.~W.}\ \bibnamefont {Nam}}, \bibinfo {author} {\bibfnamefont {G.}~\bibnamefont {Pokharel}}, \bibinfo {author} {\bibfnamefont {S.~D.}\ \bibnamefont {Wilson}}, \bibinfo {author} {\bibfnamefont {J.-H.}\ \bibnamefont {Cho}},\ and\ \bibinfo {author} {\bibfnamefont {S.~J.}\ \bibnamefont {Moon}},\ }\bibfield  {title} {\bibinfo {title} {Infrared probe of the charge density wave gap in {ScV}$_{6}${Sn}$_{6}$},\ }\href {https://doi.org/10.1103/physrevb.108.205118} {\bibfield  {journal} {\bibinfo  {journal} {Physical Review B}\ }\textbf {\bibinfo {volume} {108}},\ \bibinfo {pages} {205118} (\bibinfo {year} {2023})}\BibitemShut {NoStop}%
\bibitem [{\citenamefont {Cheng}\ \emph {et~al.}(2024{\natexlab{a}})\citenamefont {Cheng}, \citenamefont {Shao}, \citenamefont {Kim}, \citenamefont {Cochran}, \citenamefont {Yang}, \citenamefont {Yi}, \citenamefont {Jiang}, \citenamefont {Zhang}, \citenamefont {Hossain}, \citenamefont {Roychowdhury}, \citenamefont {Yilmaz}, \citenamefont {Vescovo}, \citenamefont {Fedorov}, \citenamefont {Chandra}, \citenamefont {Felser}, \citenamefont {Chang},\ and\ \citenamefont {Hasan}}]{Cheng2024_UntangleCDWBulk-Surface-ScV6Sn6}%
  \BibitemOpen
  \bibfield  {author} {\bibinfo {author} {\bibfnamefont {Z.-J.}\ \bibnamefont {Cheng}}, \bibinfo {author} {\bibfnamefont {S.}~\bibnamefont {Shao}}, \bibinfo {author} {\bibfnamefont {B.}~\bibnamefont {Kim}}, \bibinfo {author} {\bibfnamefont {T.~A.}\ \bibnamefont {Cochran}}, \bibinfo {author} {\bibfnamefont {X.~P.}\ \bibnamefont {Yang}}, \bibinfo {author} {\bibfnamefont {C.}~\bibnamefont {Yi}}, \bibinfo {author} {\bibfnamefont {Y.-X.}\ \bibnamefont {Jiang}}, \bibinfo {author} {\bibfnamefont {J.}~\bibnamefont {Zhang}}, \bibinfo {author} {\bibfnamefont {M.~S.}\ \bibnamefont {Hossain}}, \bibinfo {author} {\bibfnamefont {S.}~\bibnamefont {Roychowdhury}}, \bibinfo {author} {\bibfnamefont {T.}~\bibnamefont {Yilmaz}}, \bibinfo {author} {\bibfnamefont {E.}~\bibnamefont {Vescovo}}, \bibinfo {author} {\bibfnamefont {A.}~\bibnamefont {Fedorov}}, \bibinfo {author} {\bibfnamefont {S.}~\bibnamefont {Chandra}}, \bibinfo {author} {\bibfnamefont {C.}~\bibnamefont {Felser}}, \bibinfo {author} {\bibfnamefont {G.}~\bibnamefont
  {Chang}},\ and\ \bibinfo {author} {\bibfnamefont {M.~Z.}\ \bibnamefont {Hasan}},\ }\bibfield  {title} {\bibinfo {title} {Untangle charge-order dependent bulk states from surface effects in a topological kagome metal {ScV}$_{6}${Sn}$_{6}$}\ }\href {https://doi.org/10.48550/ARXIV.2402.02341} {10.48550/ARXIV.2402.02341} (\bibinfo {year} {2024}{\natexlab{a}}),\ \Eprint {https://arxiv.org/abs/2402.02341} {arXiv:2402.02341 [cond-mat.str-el]} \BibitemShut {NoStop}%
\bibitem [{\citenamefont {Yang}\ \emph {et~al.}(2024)\citenamefont {Yang}, \citenamefont {Cho}, \citenamefont {Li}, \citenamefont {Liu}, \citenamefont {Liu}, \citenamefont {Jiang}, \citenamefont {Ding}, \citenamefont {Xia}, \citenamefont {Tao}, \citenamefont {Liu}, \citenamefont {Jing}, \citenamefont {Huang}, \citenamefont {Shi}, \citenamefont {Huh}, \citenamefont {Kondo}, \citenamefont {Sun}, \citenamefont {Liu}, \citenamefont {Ye}, \citenamefont {Wang}, \citenamefont {Guo},\ and\ \citenamefont {Shen}}]{Yang2024_UnveilingCDWMechanism-ScV6Sn6}%
  \BibitemOpen
  \bibfield  {author} {\bibinfo {author} {\bibfnamefont {Y.-C.}\ \bibnamefont {Yang}}, \bibinfo {author} {\bibfnamefont {S.}~\bibnamefont {Cho}}, \bibinfo {author} {\bibfnamefont {T.-R.}\ \bibnamefont {Li}}, \bibinfo {author} {\bibfnamefont {X.-Q.}\ \bibnamefont {Liu}}, \bibinfo {author} {\bibfnamefont {Z.-T.}\ \bibnamefont {Liu}}, \bibinfo {author} {\bibfnamefont {Z.-C.}\ \bibnamefont {Jiang}}, \bibinfo {author} {\bibfnamefont {J.-Y.}\ \bibnamefont {Ding}}, \bibinfo {author} {\bibfnamefont {W.}~\bibnamefont {Xia}}, \bibinfo {author} {\bibfnamefont {Z.-C.}\ \bibnamefont {Tao}}, \bibinfo {author} {\bibfnamefont {J.-Y.}\ \bibnamefont {Liu}}, \bibinfo {author} {\bibfnamefont {W.-C.}\ \bibnamefont {Jing}}, \bibinfo {author} {\bibfnamefont {Y.}~\bibnamefont {Huang}}, \bibinfo {author} {\bibfnamefont {Y.-M.}\ \bibnamefont {Shi}}, \bibinfo {author} {\bibfnamefont {S.}~\bibnamefont {Huh}}, \bibinfo {author} {\bibfnamefont {T.}~\bibnamefont {Kondo}}, \bibinfo {author} {\bibfnamefont {Z.}~\bibnamefont {Sun}}, \bibinfo
  {author} {\bibfnamefont {J.-S.}\ \bibnamefont {Liu}}, \bibinfo {author} {\bibfnamefont {M.}~\bibnamefont {Ye}}, \bibinfo {author} {\bibfnamefont {Y.-L.}\ \bibnamefont {Wang}}, \bibinfo {author} {\bibfnamefont {Y.-F.}\ \bibnamefont {Guo}},\ and\ \bibinfo {author} {\bibfnamefont {D.-W.}\ \bibnamefont {Shen}},\ }\bibfield  {title} {\bibinfo {title} {Unveiling the charge density wave mechanism in vanadium-based bi-layered kagome metals}\ }\href {https://doi.org/10.48550/ARXIV.2402.03765} {10.48550/ARXIV.2402.03765} (\bibinfo {year} {2024}),\ \Eprint {https://arxiv.org/abs/2402.03765} {arXiv:2402.03765 [cond-mat.mtrl-sci]} \BibitemShut {NoStop}%
\bibitem [{\citenamefont {Kundu}\ \emph {et~al.}(2024)\citenamefont {Kundu}, \citenamefont {Huang}, \citenamefont {Seewald}, \citenamefont {Ritz}, \citenamefont {Pakhira}, \citenamefont {Zhang}, \citenamefont {Sun}, \citenamefont {Turkel}, \citenamefont {Shabani}, \citenamefont {Yilmaz}, \citenamefont {Vescovo}, \citenamefont {Dean}, \citenamefont {Johnston}, \citenamefont {Valla}, \citenamefont {Birol}, \citenamefont {Basov}, \citenamefont {Fernandes},\ and\ \citenamefont {Pasupathy}}]{Kundu2024_LowEnergyElectronicStructure-ScV6Sn6}%
  \BibitemOpen
  \bibfield  {author} {\bibinfo {author} {\bibfnamefont {A.~K.}\ \bibnamefont {Kundu}}, \bibinfo {author} {\bibfnamefont {X.}~\bibnamefont {Huang}}, \bibinfo {author} {\bibfnamefont {E.}~\bibnamefont {Seewald}}, \bibinfo {author} {\bibfnamefont {E.}~\bibnamefont {Ritz}}, \bibinfo {author} {\bibfnamefont {S.}~\bibnamefont {Pakhira}}, \bibinfo {author} {\bibfnamefont {S.}~\bibnamefont {Zhang}}, \bibinfo {author} {\bibfnamefont {D.}~\bibnamefont {Sun}}, \bibinfo {author} {\bibfnamefont {S.}~\bibnamefont {Turkel}}, \bibinfo {author} {\bibfnamefont {S.}~\bibnamefont {Shabani}}, \bibinfo {author} {\bibfnamefont {T.}~\bibnamefont {Yilmaz}}, \bibinfo {author} {\bibfnamefont {E.}~\bibnamefont {Vescovo}}, \bibinfo {author} {\bibfnamefont {C.~R.}\ \bibnamefont {Dean}}, \bibinfo {author} {\bibfnamefont {D.~C.}\ \bibnamefont {Johnston}}, \bibinfo {author} {\bibfnamefont {T.}~\bibnamefont {Valla}}, \bibinfo {author} {\bibfnamefont {T.}~\bibnamefont {Birol}}, \bibinfo {author} {\bibfnamefont {D.~N.}\ \bibnamefont {Basov}},
  \bibinfo {author} {\bibfnamefont {R.~M.}\ \bibnamefont {Fernandes}},\ and\ \bibinfo {author} {\bibfnamefont {A.~N.}\ \bibnamefont {Pasupathy}},\ }\bibfield  {title} {\bibinfo {title} {Low-energy electronic structure in the unconventional charge-ordered state of {ScV}$_{6}${Sn}$_{6}$},\ }\bibfield  {journal} {\bibinfo  {journal} {Nature Communications}\ }\textbf {\bibinfo {volume} {15}},\ \href {https://doi.org/10.1038/s41467-024-48883-0} {10.1038/s41467-024-48883-0} (\bibinfo {year} {2024})\BibitemShut {NoStop}%
\bibitem [{\citenamefont {Zhang}\ \emph {et~al.}(2022)\citenamefont {Zhang}, \citenamefont {Hou}, \citenamefont {Xia}, \citenamefont {Xu}, \citenamefont {Yang}, \citenamefont {Wang}, \citenamefont {Liu}, \citenamefont {Shen}, \citenamefont {Zhang}, \citenamefont {Dong}, \citenamefont {Uwatoko}, \citenamefont {Sun}, \citenamefont {Wang}, \citenamefont {Guo},\ and\ \citenamefont {Cheng}}]{Zhang2022_ScV6Sn6-Pressure}%
  \BibitemOpen
  \bibfield  {author} {\bibinfo {author} {\bibfnamefont {X.}~\bibnamefont {Zhang}}, \bibinfo {author} {\bibfnamefont {J.}~\bibnamefont {Hou}}, \bibinfo {author} {\bibfnamefont {W.}~\bibnamefont {Xia}}, \bibinfo {author} {\bibfnamefont {Z.}~\bibnamefont {Xu}}, \bibinfo {author} {\bibfnamefont {P.}~\bibnamefont {Yang}}, \bibinfo {author} {\bibfnamefont {A.}~\bibnamefont {Wang}}, \bibinfo {author} {\bibfnamefont {Z.}~\bibnamefont {Liu}}, \bibinfo {author} {\bibfnamefont {J.}~\bibnamefont {Shen}}, \bibinfo {author} {\bibfnamefont {H.}~\bibnamefont {Zhang}}, \bibinfo {author} {\bibfnamefont {X.}~\bibnamefont {Dong}}, \bibinfo {author} {\bibfnamefont {Y.}~\bibnamefont {Uwatoko}}, \bibinfo {author} {\bibfnamefont {J.}~\bibnamefont {Sun}}, \bibinfo {author} {\bibfnamefont {B.}~\bibnamefont {Wang}}, \bibinfo {author} {\bibfnamefont {Y.}~\bibnamefont {Guo}},\ and\ \bibinfo {author} {\bibfnamefont {J.}~\bibnamefont {Cheng}},\ }\bibfield  {title} {\bibinfo {title} {Destabilization of the charge density wave and the
  absence of superconductivity in {ScV}$_{6}${Sn}$_{6}$ under high pressures up to 11 {GPa}},\ }\href {https://doi.org/10.3390/ma15207372} {\bibfield  {journal} {\bibinfo  {journal} {Materials}\ }\textbf {\bibinfo {volume} {15}},\ \bibinfo {pages} {7372} (\bibinfo {year} {2022})}\BibitemShut {NoStop}%
\bibitem [{\citenamefont {Meier}\ \emph {et~al.}(2023)\citenamefont {Meier}, \citenamefont {Madhogaria}, \citenamefont {Mozaffari}, \citenamefont {Marshall}, \citenamefont {Graf}, \citenamefont {McGuire}, \citenamefont {Arachchige}, \citenamefont {Allen}, \citenamefont {Driver}, \citenamefont {Cao},\ and\ \citenamefont {Mandrus}}]{Meier2023_TinyScAllowChainsToRattle-Lu+YDopedScV6Sn6}%
  \BibitemOpen
  \bibfield  {author} {\bibinfo {author} {\bibfnamefont {W.~R.}\ \bibnamefont {Meier}}, \bibinfo {author} {\bibfnamefont {R.~P.}\ \bibnamefont {Madhogaria}}, \bibinfo {author} {\bibfnamefont {S.}~\bibnamefont {Mozaffari}}, \bibinfo {author} {\bibfnamefont {M.}~\bibnamefont {Marshall}}, \bibinfo {author} {\bibfnamefont {D.~E.}\ \bibnamefont {Graf}}, \bibinfo {author} {\bibfnamefont {M.~A.}\ \bibnamefont {McGuire}}, \bibinfo {author} {\bibfnamefont {H.~W.~S.}\ \bibnamefont {Arachchige}}, \bibinfo {author} {\bibfnamefont {C.~L.}\ \bibnamefont {Allen}}, \bibinfo {author} {\bibfnamefont {J.}~\bibnamefont {Driver}}, \bibinfo {author} {\bibfnamefont {H.}~\bibnamefont {Cao}},\ and\ \bibinfo {author} {\bibfnamefont {D.}~\bibnamefont {Mandrus}},\ }\bibfield  {title} {\bibinfo {title} {Tiny {Sc} allows the chains to rattle: Impact of {Lu} and {Y} doping on the charge-density wave in {ScV}$_{6}${Sn}$_{6}$},\ }\href {https://doi.org/10.1021/jacs.3c06394} {\bibfield  {journal} {\bibinfo  {journal} {Journal of the American
  Chemical Society}\ }\textbf {\bibinfo {volume} {145}},\ \bibinfo {pages} {20943} (\bibinfo {year} {2023})}\BibitemShut {NoStop}%
\bibitem [{\citenamefont {Cheng}\ \emph {et~al.}(2024{\natexlab{b}})\citenamefont {Cheng}, \citenamefont {Ren}, \citenamefont {Li}, \citenamefont {Oh}, \citenamefont {Tan}, \citenamefont {Pokharel}, \citenamefont {DeStefano}, \citenamefont {Rosenberg}, \citenamefont {Guo}, \citenamefont {Zhang}, \citenamefont {Yue}, \citenamefont {Lee}, \citenamefont {Gorovikov}, \citenamefont {Zonno}, \citenamefont {Hashimoto}, \citenamefont {Lu}, \citenamefont {Ke}, \citenamefont {Mazzola}, \citenamefont {Kono}, \citenamefont {Birgeneau}, \citenamefont {Chu}, \citenamefont {Wilson}, \citenamefont {Wang}, \citenamefont {Yan}, \citenamefont {Yi},\ and\ \citenamefont {Zeljkovic}}]{Cheng2024_NanoscaleVisualizationOfCDW-ScV6Sn6}%
  \BibitemOpen
  \bibfield  {author} {\bibinfo {author} {\bibfnamefont {S.}~\bibnamefont {Cheng}}, \bibinfo {author} {\bibfnamefont {Z.}~\bibnamefont {Ren}}, \bibinfo {author} {\bibfnamefont {H.}~\bibnamefont {Li}}, \bibinfo {author} {\bibfnamefont {J.~S.}\ \bibnamefont {Oh}}, \bibinfo {author} {\bibfnamefont {H.}~\bibnamefont {Tan}}, \bibinfo {author} {\bibfnamefont {G.}~\bibnamefont {Pokharel}}, \bibinfo {author} {\bibfnamefont {J.~M.}\ \bibnamefont {DeStefano}}, \bibinfo {author} {\bibfnamefont {E.}~\bibnamefont {Rosenberg}}, \bibinfo {author} {\bibfnamefont {Y.}~\bibnamefont {Guo}}, \bibinfo {author} {\bibfnamefont {Y.}~\bibnamefont {Zhang}}, \bibinfo {author} {\bibfnamefont {Z.}~\bibnamefont {Yue}}, \bibinfo {author} {\bibfnamefont {Y.}~\bibnamefont {Lee}}, \bibinfo {author} {\bibfnamefont {S.}~\bibnamefont {Gorovikov}}, \bibinfo {author} {\bibfnamefont {M.}~\bibnamefont {Zonno}}, \bibinfo {author} {\bibfnamefont {M.}~\bibnamefont {Hashimoto}}, \bibinfo {author} {\bibfnamefont {D.}~\bibnamefont {Lu}}, \bibinfo {author}
  {\bibfnamefont {L.}~\bibnamefont {Ke}}, \bibinfo {author} {\bibfnamefont {F.}~\bibnamefont {Mazzola}}, \bibinfo {author} {\bibfnamefont {J.}~\bibnamefont {Kono}}, \bibinfo {author} {\bibfnamefont {R.~J.}\ \bibnamefont {Birgeneau}}, \bibinfo {author} {\bibfnamefont {J.-H.}\ \bibnamefont {Chu}}, \bibinfo {author} {\bibfnamefont {S.~D.}\ \bibnamefont {Wilson}}, \bibinfo {author} {\bibfnamefont {Z.}~\bibnamefont {Wang}}, \bibinfo {author} {\bibfnamefont {B.}~\bibnamefont {Yan}}, \bibinfo {author} {\bibfnamefont {M.}~\bibnamefont {Yi}},\ and\ \bibinfo {author} {\bibfnamefont {I.}~\bibnamefont {Zeljkovic}},\ }\bibfield  {title} {\bibinfo {title} {Nanoscale visualization and spectral fingerprints of the charge order in {ScV}$_{6}${Sn}$_{6}$ distinct from other kagome metals},\ }\bibfield  {journal} {\bibinfo  {journal} {npj Quantum Materials}\ }\textbf {\bibinfo {volume} {9}},\ \href {https://doi.org/10.1038/s41535-024-00623-9} {10.1038/s41535-024-00623-9} (\bibinfo {year} {2024}{\natexlab{b}})\BibitemShut
  {NoStop}%
\bibitem [{\citenamefont {Hu}\ \emph {et~al.}(2024)\citenamefont {Hu}, \citenamefont {Ma}, \citenamefont {Li}, \citenamefont {Jiang}, \citenamefont {Gawryluk}, \citenamefont {Hu}, \citenamefont {Teyssier}, \citenamefont {Multian}, \citenamefont {Yin}, \citenamefont {Xu}, \citenamefont {Shin}, \citenamefont {Plokhikh}, \citenamefont {Han}, \citenamefont {Plumb}, \citenamefont {Liu}, \citenamefont {Yin}, \citenamefont {Guguchia}, \citenamefont {Zhao}, \citenamefont {Schnyder}, \citenamefont {Wu}, \citenamefont {Pomjakushina}, \citenamefont {Hasan}, \citenamefont {Wang},\ and\ \citenamefont {Shi}}]{Hu2024_PhononPromotedCDW-ScV6Sn6}%
  \BibitemOpen
  \bibfield  {author} {\bibinfo {author} {\bibfnamefont {Y.}~\bibnamefont {Hu}}, \bibinfo {author} {\bibfnamefont {J.}~\bibnamefont {Ma}}, \bibinfo {author} {\bibfnamefont {Y.}~\bibnamefont {Li}}, \bibinfo {author} {\bibfnamefont {Y.}~\bibnamefont {Jiang}}, \bibinfo {author} {\bibfnamefont {D.~J.}\ \bibnamefont {Gawryluk}}, \bibinfo {author} {\bibfnamefont {T.}~\bibnamefont {Hu}}, \bibinfo {author} {\bibfnamefont {J.}~\bibnamefont {Teyssier}}, \bibinfo {author} {\bibfnamefont {V.}~\bibnamefont {Multian}}, \bibinfo {author} {\bibfnamefont {Z.}~\bibnamefont {Yin}}, \bibinfo {author} {\bibfnamefont {S.}~\bibnamefont {Xu}}, \bibinfo {author} {\bibfnamefont {S.}~\bibnamefont {Shin}}, \bibinfo {author} {\bibfnamefont {I.}~\bibnamefont {Plokhikh}}, \bibinfo {author} {\bibfnamefont {X.}~\bibnamefont {Han}}, \bibinfo {author} {\bibfnamefont {N.~C.}\ \bibnamefont {Plumb}}, \bibinfo {author} {\bibfnamefont {Y.}~\bibnamefont {Liu}}, \bibinfo {author} {\bibfnamefont {J.-X.}\ \bibnamefont {Yin}}, \bibinfo {author}
  {\bibfnamefont {Z.}~\bibnamefont {Guguchia}}, \bibinfo {author} {\bibfnamefont {Y.}~\bibnamefont {Zhao}}, \bibinfo {author} {\bibfnamefont {A.~P.}\ \bibnamefont {Schnyder}}, \bibinfo {author} {\bibfnamefont {X.}~\bibnamefont {Wu}}, \bibinfo {author} {\bibfnamefont {E.}~\bibnamefont {Pomjakushina}}, \bibinfo {author} {\bibfnamefont {M.~Z.}\ \bibnamefont {Hasan}}, \bibinfo {author} {\bibfnamefont {N.}~\bibnamefont {Wang}},\ and\ \bibinfo {author} {\bibfnamefont {M.}~\bibnamefont {Shi}},\ }\bibfield  {title} {\bibinfo {title} {Phonon promoted charge density wave in topological kagome metal {ScV}$_{6}${Sn}$_{6}$},\ }\bibfield  {journal} {\bibinfo  {journal} {Nature Communications}\ }\textbf {\bibinfo {volume} {15}},\ \href {https://doi.org/10.1038/s41467-024-45859-y} {10.1038/s41467-024-45859-y} (\bibinfo {year} {2024})\BibitemShut {NoStop}%
\bibitem [{\citenamefont {Ortiz}\ \emph {et~al.}(2025)\citenamefont {Ortiz}, \citenamefont {Meier}, \citenamefont {Pokharel}, \citenamefont {Chamorro}, \citenamefont {Yang}, \citenamefont {Mozaffari}, \citenamefont {Thaler}, \citenamefont {Gomez~Alvarado}, \citenamefont {Zhang}, \citenamefont {Parker}, \citenamefont {Samolyuk}, \citenamefont {Paddison}, \citenamefont {Yan}, \citenamefont {Ye}, \citenamefont {Sarker}, \citenamefont {Wilson}, \citenamefont {Miao}, \citenamefont {Mandrus},\ and\ \citenamefont {McGuire}}]{Ortiz2025_166StabilityFrontiers+LnNb6Sn6-Family+DW-LuNb6Sn6}%
  \BibitemOpen
  \bibfield  {author} {\bibinfo {author} {\bibfnamefont {B.~R.}\ \bibnamefont {Ortiz}}, \bibinfo {author} {\bibfnamefont {W.~R.}\ \bibnamefont {Meier}}, \bibinfo {author} {\bibfnamefont {G.}~\bibnamefont {Pokharel}}, \bibinfo {author} {\bibfnamefont {J.}~\bibnamefont {Chamorro}}, \bibinfo {author} {\bibfnamefont {F.}~\bibnamefont {Yang}}, \bibinfo {author} {\bibfnamefont {S.}~\bibnamefont {Mozaffari}}, \bibinfo {author} {\bibfnamefont {A.}~\bibnamefont {Thaler}}, \bibinfo {author} {\bibfnamefont {S.~J.}\ \bibnamefont {Gomez~Alvarado}}, \bibinfo {author} {\bibfnamefont {H.}~\bibnamefont {Zhang}}, \bibinfo {author} {\bibfnamefont {D.~S.}\ \bibnamefont {Parker}}, \bibinfo {author} {\bibfnamefont {G.~D.}\ \bibnamefont {Samolyuk}}, \bibinfo {author} {\bibfnamefont {J.~A.~M.}\ \bibnamefont {Paddison}}, \bibinfo {author} {\bibfnamefont {J.}~\bibnamefont {Yan}}, \bibinfo {author} {\bibfnamefont {F.}~\bibnamefont {Ye}}, \bibinfo {author} {\bibfnamefont {S.}~\bibnamefont {Sarker}}, \bibinfo {author} {\bibfnamefont
  {S.~D.}\ \bibnamefont {Wilson}}, \bibinfo {author} {\bibfnamefont {H.}~\bibnamefont {Miao}}, \bibinfo {author} {\bibfnamefont {D.}~\bibnamefont {Mandrus}},\ and\ \bibinfo {author} {\bibfnamefont {M.~A.}\ \bibnamefont {McGuire}},\ }\bibfield  {title} {\bibinfo {title} {Stability frontiers in the $am_{6}x_{6}$ {K}agome metals: The $ln${Nb}$_{6}${Sn}$_{6}$ ($ln$:{Ce}-{Lu},{Y}) family and density-wave transition in {LuNb}$_6${Sn}$_6$},\ }\href {https://doi.org/10.1021/jacs.4c16347} {\bibfield  {journal} {\bibinfo  {journal} {Journal of the American Chemical Society}\ }\textbf {\bibinfo {volume} {147}},\ \bibinfo {pages} {5279} (\bibinfo {year} {2025})}\BibitemShut {NoStop}%
\bibitem [{\citenamefont {Canfield}\ \emph {et~al.}(2016)\citenamefont {Canfield}, \citenamefont {Kong}, \citenamefont {Kaluarachchi},\ and\ \citenamefont {Jo}}]{Canfield2016_CanfieldCrucibleSets}%
  \BibitemOpen
  \bibfield  {author} {\bibinfo {author} {\bibfnamefont {P.~C.}\ \bibnamefont {Canfield}}, \bibinfo {author} {\bibfnamefont {T.}~\bibnamefont {Kong}}, \bibinfo {author} {\bibfnamefont {U.~S.}\ \bibnamefont {Kaluarachchi}},\ and\ \bibinfo {author} {\bibfnamefont {N.~H.}\ \bibnamefont {Jo}},\ }\bibfield  {title} {\bibinfo {title} {Use of frit-disc crucibles for routine and exploratory solution growth of single crystalline samples},\ }\href {https://doi.org/10.1080/14786435.2015.1122248} {\bibfield  {journal} {\bibinfo  {journal} {Philos. Mag.}\ }\textbf {\bibinfo {volume} {96}},\ \bibinfo {pages} {84} (\bibinfo {year} {2016})}\BibitemShut {NoStop}%
\bibitem [{\citenamefont {Staško}\ \emph {et~al.}(2020)\citenamefont {Staško}, \citenamefont {Prchal}, \citenamefont {Klicpera}, \citenamefont {Aoki},\ and\ \citenamefont {Murata}}]{Stasko2020_PressureMediaDaphneOil7000}%
  \BibitemOpen
  \bibfield  {author} {\bibinfo {author} {\bibfnamefont {D.}~\bibnamefont {Staško}}, \bibinfo {author} {\bibfnamefont {J.}~\bibnamefont {Prchal}}, \bibinfo {author} {\bibfnamefont {M.}~\bibnamefont {Klicpera}}, \bibinfo {author} {\bibfnamefont {S.}~\bibnamefont {Aoki}},\ and\ \bibinfo {author} {\bibfnamefont {K.}~\bibnamefont {Murata}},\ }\bibfield  {title} {\bibinfo {title} {Pressure media for high pressure experiments, {D}aphne oil 7000 series},\ }\href {https://doi.org/10.1080/08957959.2020.1825706} {\bibfield  {journal} {\bibinfo  {journal} {High Pressure Research}\ }\textbf {\bibinfo {volume} {40}},\ \bibinfo {pages} {525} (\bibinfo {year} {2020})}\BibitemShut {NoStop}%
\bibitem [{LuN(2025)}]{LuNb6Sn6Pressure_SupplementaryData}%
  \BibitemOpen
  \href@noop {} {\bibinfo {title} {See supplemental material at [url] for resistance data.}} (\bibinfo {year} {2025})\BibitemShut {NoStop}%
\bibitem [{\citenamefont {Piermarini}\ \emph {et~al.}(1975)\citenamefont {Piermarini}, \citenamefont {Block}, \citenamefont {Barnett},\ and\ \citenamefont {Forman}}]{Piermarini1975_RubyFluorescenceTo195kbar}%
  \BibitemOpen
  \bibfield  {author} {\bibinfo {author} {\bibfnamefont {G.~J.}\ \bibnamefont {Piermarini}}, \bibinfo {author} {\bibfnamefont {S.}~\bibnamefont {Block}}, \bibinfo {author} {\bibfnamefont {J.~D.}\ \bibnamefont {Barnett}},\ and\ \bibinfo {author} {\bibfnamefont {R.~A.}\ \bibnamefont {Forman}},\ }\bibfield  {title} {\bibinfo {title} {Calibration of the pressure dependence of the ${R}_{1}$ ruby fluorescence line to 195 kbar},\ }\href {https://doi.org/10.1063/1.321957} {\bibfield  {journal} {\bibinfo  {journal} {Journal of Applied Physics}\ }\textbf {\bibinfo {volume} {46}},\ \bibinfo {pages} {2774} (\bibinfo {year} {1975})}\BibitemShut {NoStop}%
\bibitem [{\citenamefont {Hu}\ \emph {et~al.}(2023)\citenamefont {Hu}, \citenamefont {Pi}, \citenamefont {Xu}, \citenamefont {Yue}, \citenamefont {Wu}, \citenamefont {Liu}, \citenamefont {Zhang}, \citenamefont {Li}, \citenamefont {Zhou}, \citenamefont {Yuan}, \citenamefont {Wu}, \citenamefont {Dong}, \citenamefont {Weng},\ and\ \citenamefont {Wang}}]{Hu2023_OpticalConductivity+BandStructure-ScV6Sn6}%
  \BibitemOpen
  \bibfield  {author} {\bibinfo {author} {\bibfnamefont {T.}~\bibnamefont {Hu}}, \bibinfo {author} {\bibfnamefont {H.}~\bibnamefont {Pi}}, \bibinfo {author} {\bibfnamefont {S.}~\bibnamefont {Xu}}, \bibinfo {author} {\bibfnamefont {L.}~\bibnamefont {Yue}}, \bibinfo {author} {\bibfnamefont {Q.}~\bibnamefont {Wu}}, \bibinfo {author} {\bibfnamefont {Q.}~\bibnamefont {Liu}}, \bibinfo {author} {\bibfnamefont {S.}~\bibnamefont {Zhang}}, \bibinfo {author} {\bibfnamefont {R.}~\bibnamefont {Li}}, \bibinfo {author} {\bibfnamefont {X.}~\bibnamefont {Zhou}}, \bibinfo {author} {\bibfnamefont {J.}~\bibnamefont {Yuan}}, \bibinfo {author} {\bibfnamefont {D.}~\bibnamefont {Wu}}, \bibinfo {author} {\bibfnamefont {T.}~\bibnamefont {Dong}}, \bibinfo {author} {\bibfnamefont {H.}~\bibnamefont {Weng}},\ and\ \bibinfo {author} {\bibfnamefont {N.}~\bibnamefont {Wang}},\ }\bibfield  {title} {\bibinfo {title} {Optical spectroscopy and band structure calculations of the structural phase transition in the vanadium-based kagome metal
  {ScV}$_{6}${Sn}$_{6}$},\ }\href {https://doi.org/10.1103/physrevb.107.165119} {\bibfield  {journal} {\bibinfo  {journal} {Physical Review B}\ }\textbf {\bibinfo {volume} {107}},\ \bibinfo {pages} {165119} (\bibinfo {year} {2023})}\BibitemShut {NoStop}%
\bibitem [{\citenamefont {Brandt}\ and\ \citenamefont {Ginzburg}(1965)}]{Brandt1965_SuperconductingMetalsUnderPressure}%
  \BibitemOpen
  \bibfield  {author} {\bibinfo {author} {\bibfnamefont {N.~B.}\ \bibnamefont {Brandt}}\ and\ \bibinfo {author} {\bibfnamefont {N.~I.}\ \bibnamefont {Ginzburg}},\ }\bibfield  {title} {\bibinfo {title} {Effect of high pressure on the superconducting properties of metals},\ }\href {https://doi.org/10.1070/pu1965v008n02abeh003037} {\bibfield  {journal} {\bibinfo  {journal} {Soviet Physics Uspekhi}\ }\textbf {\bibinfo {volume} {8}},\ \bibinfo {pages} {202} (\bibinfo {year} {1965})}\BibitemShut {NoStop}%
\bibitem [{\citenamefont {Gr\"uner}(1994)}]{Gruner1994_DensityWavesInSolids}%
  \BibitemOpen
  \bibfield  {author} {\bibinfo {author} {\bibfnamefont {G.}~\bibnamefont {Gr\"uner}},\ }\href@noop {} {\emph {\bibinfo {title} {Density waves in solids}}}\ (\bibinfo  {publisher} {Perseus Publishing, Cambridge, Mass.},\ \bibinfo {year} {1994})\BibitemShut {NoStop}%
\bibitem [{\citenamefont {Liu}\ \emph {et~al.}(2024)\citenamefont {Liu}, \citenamefont {Wang}, \citenamefont {Yao}, \citenamefont {Jia}, \citenamefont {Zhang},\ and\ \citenamefont {Cho}}]{Liu2024_DrivingMechanism+FluctuationsOfCDWs-ScV6Sn6}%
  \BibitemOpen
  \bibfield  {author} {\bibinfo {author} {\bibfnamefont {S.}~\bibnamefont {Liu}}, \bibinfo {author} {\bibfnamefont {C.}~\bibnamefont {Wang}}, \bibinfo {author} {\bibfnamefont {S.}~\bibnamefont {Yao}}, \bibinfo {author} {\bibfnamefont {Y.}~\bibnamefont {Jia}}, \bibinfo {author} {\bibfnamefont {Z.}~\bibnamefont {Zhang}},\ and\ \bibinfo {author} {\bibfnamefont {J.-H.}\ \bibnamefont {Cho}},\ }\bibfield  {title} {\bibinfo {title} {Driving mechanism and dynamic fluctuations of charge density waves in the kagome metal {ScV}$_{6}${Sn}$_{6}$},\ }\href {https://doi.org/10.1103/physrevb.109.l121103} {\bibfield  {journal} {\bibinfo  {journal} {Physical Review B}\ }\textbf {\bibinfo {volume} {109}},\ \bibinfo {pages} {l121103} (\bibinfo {year} {2024})}\BibitemShut {NoStop}%
\end{thebibliography}

%

\end{document}